\documentclass[preprintnumbers,prd,showpacs,floatfix,superscriptaddress,nofootinbib,twocolumn]{revtex4-1}

\pdfoutput=1
\usepackage{graphicx,epsfig}
\usepackage{amsmath}
\usepackage{amssymb}
\usepackage{longtable}
\usepackage{multirow}
\usepackage{dcolumn}
\usepackage{bm}
\usepackage{amsfonts}
\usepackage{subfigure}
\usepackage{color}
\usepackage{relsize}
\usepackage{url}
\usepackage{soul}
\usepackage{xcolor}
\usepackage[linktocpage]{hyperref}

\def\be{\begin{equation}}
\def\ee{\end{equation}}
\newcommand{\beq}{\begin{eqnarray}}
\newcommand{\eeq}{\end{eqnarray}}

\begin{document}

\title{Observational signatures of hot spots orbiting horizonless objects}
%
%
\author{Jo\~{a}o Lu\'{i}s Rosa}
\email{joaoluis92@gmail.com}
\affiliation{Institute of Physics, University of Tartu, W. Ostwaldi 1, 50411 Tartu, Estonia}
\author{Paulo Garcia}
\email{pgarcia@fe.up.pt}
\affiliation{Faculdade de Engenharia, Universidade do Porto, rua Dr. Roberto Frias, 4200-465 Porto, Portugal}
\affiliation{CENTRA, Departamento de F\'{i}sica, Instituto Superior T\'{e}cnico -- IST, Universidade de Lisboa -- UL, Avenida Rovisco Pais 1, 1049-001 Lisboa, Portugal}
\author{Fr\'{e}d\'{e}ric H. Vincent}
\affiliation{LESIA, Observatoire de Paris, Universit\'{e} PSL, CNRS, Sorbonne Universit\'{e},
Universit\'{e} de Paris, 5 place Jules Janssen, 92190 Meudon, France}
\author{Vitor Cardoso}
\affiliation{CENTRA, Departamento de F\'{i}sica, Instituto Superior T\'{e}cnico -- IST, Universidade de Lisboa -- UL, Avenida Rovisco Pais 1, 1049-001 Lisboa, Portugal}
\affiliation{Niels Bohr International Academy, Niels Bohr Institute, Blegdamsvej 17, 2100 Copenhagen, Denmark}
\date{\today}

\begin{abstract} 
Pushed by a number of advances, electromagnetic observatories have now reached the horizon scale of supermassive black holes. The existence and properties of horizons in our universe is one of the outstanding fundamental issues that can now be addressed.
Here we investigate the ability to discriminate between black holes and compact, horizonless objects, focusing on the lensing of hot spots around compact objects. We work in particular with boson and Proca stars as central objects, and show that the absence of a horizon gives rise to a characteristic feature -- photons that plough through the central object and produce an extra image. This feature should be universal for central objects made of matter weakly coupled to the standard model.

\end{abstract}

\maketitle

\section{Introduction}\label{sec:intro}
Monitoring of regions of strong-field gravity is now possible with a wide array of observatories, including gravitational-wave detectors~\cite{Abbott:2016blz,Abbott:2020niy}
and very long baseline interferometry~\cite{Akiyama:2019cqa,Abuter:2020dou}. These observatories will be updated in the years to come, and others will be added to the network,
giving us access to clean data regarding the behaviour of gravity
in the most extreme circumstances. Observations regarding gravitational fields at its extreme can inform us on some of the outstanding issues regarding the gravitational interaction~\cite{Barack:2018yly,Cardoso:2019rvt}: are observations consistent with the uniqueness results of General Relativity, according to which isolated black holes (BHs) all belong to the same family of solutions -- the Kerr family~\cite{Kerr:1963ud} -- fully described by two parameters alone, mass and angular momentum~\cite{Chrusciel:2012jk,Robinson:2004zz}. In fact, are observations consistent with the BH paradigm,
do BHs exist~\cite{Barack:2018yly,Cardoso:2019rvt}?

We focus here on tests of the nature of the dark massive objects found at the centre of most galaxies, via electromagnetic observations. There is ample theoretical support for these being BHs, regions of spacetime which are causally disconnected from ours via a one-way membrane, the horizon. In fact, there is no known mechanism to prevent massive stars from eventually collapsing to BHs. However, the collapse of ``reasonable'' matter always leads to the formation of singularities, regions where the description of the gravitational interaction breaks down~\cite{Penrose:1964wq,Penrose:1969pc}. It is thus an extraordinary statement that the universe produces BHs, and that any of these hides and shields us from the failure of the theory from which they derive. Among others, such remarkable property certainly deserves observational scrutiny~\cite{Cardoso:2019rvt}.

Now, to collect evidence of the existence of BHs, one needs to perform observations of phenomena close to the horizon. Photons which are emitted from such regions, directed into the BH and with a small enough impact parameter are simply absorbed by the BH. It follows that BHs cast a shadow~\cite{1972ApJ...173L.137C,Luminet:1979nyg,Falcke:1999pj,Cardoso:2019dte,Gralla:2020srx,Cunha:2018acu,Cardoso:2021sip}, the geometry of which depends on the assumptions about the accretion flow~\cite{Vincent:2022fwj}.
This has been a field with intense activity lately, mostly focusing on producing images of accreting BHs. It turns out (not too surprisingly) that other compact objects are equally able to cast shadows which to a good precision looks similar to that of BHs~\cite{Vincent:2015xta,Cardoso:2019rvt,Rosa:2022tfv}. Here we wish to focus our attention not on accretion disks but on localised emitting sources, which we will call hot spots. These are thought to arise in reconnecting events in a magnetised accretion disk \cite{ 2021arXiv210708056S, 2020MNRAS.497.4999D,elmellah21} and have been frequently observed in the center of our own galaxy \cite{2021ApJ...917...73W, 2021ApJ...923...54M, 2021AA...654A..22G}. An exciting aspect of these hot spots is that their motion in the plane of the sky was detected and found to be compatible with orbital motion near the innermost circular last stable orbit \cite{Gravity2018,Baubock:2020dgq}. They hold the promise to be an interesting tool to discriminate between BHs and other, horizonless, compact objects.


\section{Theory and framework}
There are different strategies to test the nature of supermassive compact objects, specially in regimes appropriate to many electromagnetic campaigns, where the spacetime is mostly stationary. One is to simply write down a parametrized metric which satisfies some basic requirements, such as asymptotic flatness. These spacetimes invariably have a matter content which is not physically motivated and may even violate some of our most cherished energy conditions. An alternative, which we follow here, is to focus on certain matter contents and work out the geometry from the field equations. 

We consider the simplest possible matter content, a massive bosonic field ( scalar or a vector), minimally coupled to gravity. We briefly describe the two different matter sectors and their self-gravitating solutions below. 
\subsection{The Einstein-Klein-Gordon theory}\label{sec:setup}

\subsubsection{Action and equations of motion}
We start with a theory describing a massive and complex scalar field $\Phi$ minimally coupled to gravity. The action $S$ for this so-called ``Einstein Klein-Gordon'' theory is written in the form 
\begin{equation}
S=\int_\Omega\sqrt{-g}\left(\frac{R}{16\pi}-\nabla_a\bar\Phi\nabla^a\Phi-\mu^2\bar\Phi\Phi\right)d^4x\,,\label{action}
\end{equation}
where $\Omega$ is the spacetime manifold on which a metric $g_{ab}$ is defined, the metric determinant is $g$, $R=g^{ab}R_{ab}$ is the Ricci scalar, with $R_{ab}$ the Ricci tensor, and $\nabla_a$ denotes covariant derivatives written in terms of the metric $g_{ab}$. The scalar is complex and the overbar $(\bar{\ })$ denotes complex conjugation. 
We use geometrized units in this section, such that $G=c=1$, where $G$ is the gravitational constant and $c$ is the speed of light. 
In the action (\ref{action}), $\mu$ is a mass parameter representing the mass of the scalar field $\Phi$. It is related to a physical mass $m_s$ via Planck constant, $m_s=\hbar \mu$.

The theory above is interesting as a toy model, but possibly also an accurate description of dark matter, or any new fundamental bosonic field. The gravitational interaction together with the intrinsic pressure allow self-gravitating equilibrium solutions to exist. Self-gravitating solutions for the theory above are broadly referred to as boson stars,  and can be generalized through the inclusion of non-linear self-interactions~\cite{Kaup:1968zz,Ruffini:1969qy,Khlopov:1985jw,Seidel:1991zh,Guth:2014hsa,Brito:2015pxa,Minamitsuji:2018kof} (see Refs.~\cite{Jetzer:1991jr,Schunck:2003kk,Liebling:2012fv,Macedo:2013jja,Cardoso:2019rvt} for reviews).  
If the scalar field is {\it complex}, there are {\it static}, spherically-symmetric geometries, while the field itself oscillates~\cite{Kaup:1968zz,Ruffini:1969qy} (for 
reviews, see Refs.~\cite{Jetzer:1991jr,Schunck:2003kk,Liebling:2012fv,Macedo:2013jja}). Analogous solutions for complex massive vector fields were also shown to exist~\cite{Brito:2015pxa}, and we dwell on these in the next subsection. Boson stars can be compact, with gravitational potential $U ={\cal O}(0.1)$ for the simplest model, and higher when self interactions are considered.

The theory is controlled by the dimensionless coupling
\begin{equation}
\frac{G}{c\hbar} M\mu = 7.5\cdot 10^{9} \left(\frac{M}{M_{\odot}}\right) 
\left(\frac{m_sc^2}{\rm eV}\right)\,,\label{dimensionless_massparameter}
\end{equation}
where $M$ is the total mass of the boson star. 

The equilibrium solutions for this framework are obtained by solving the equations of motion. As the action in Eq.\eqref{action} depends on two independent quantities, the metric $g_{ab}$ and the scalar field $\Phi$, these equations can be derived by taking the variation of Eq.~\eqref{action} with respect to $g_{ab}$ and $\Phi$, yielding 
\beq
R_{ab}-\frac{1}{2}g_{ab}R&=&8\pi T_{ab}\,,\label{field}\\
T_{ab}&=&\frac{1}{2}\left(\nabla_a\bar\Phi\nabla_b\Phi+\nabla_a\Phi\nabla_b\bar\Phi\right)\nonumber\\
&-&\frac{1}{2}g_{ab}\left(\nabla_c\bar\Phi\nabla^c\Phi+\mu^2\bar\Phi\Phi\right)\,,\label{setensor}\\
\left(\Box-\mu^2\right)\Phi&=&0\,,\label{kleingordon}
\eeq
respectively, where $T_{ab}$ is the stress-energy tensor and $\Box=\nabla_a\nabla^a$ is the d'Alembert operator. 

\subsubsection{Equilibrium configurations: boson stars}\label{sec:bs}

For simplicity, we focus exclusively on spherically symmetric solutions of the Einstein-Klein-Gordon system of Eqs.\eqref{field}-\eqref{kleingordon}. Thus, consider the general spherically symmetric metric $g_{ab}$ described by the line element written in the usual spherical coordinates $\left(t,r,\theta,\phi\right)$ as
%
%
\begin{equation}
ds^2=-\sigma^2Ndt^2+\frac{dr^2}{N}+r^2\left(d\theta^2+\sin^2\theta d\phi^2\right)\,,\label{metric2}
\end{equation}
where $\sigma=\sigma\left(r\right)$ is an arbitrary function of $r$ and $N=N\left(r\right)$ is defined as $N\left(r\right)=1-2m\left(r\right)/r$, where $m\left(r\right)$ is a function that plays the role of the spacetime mass.
The ansatz above assumes staticity of the geometry already. 
To preserve the time independence of the metric $g_{ab}$ and the stress-energy tensor $T_{ab}$ one makes use of the $U\left(1\right)$ symmetry of Eq.\eqref{action}. To do so, consider a standing wave \textit{ansatz} for the scalar field,
\begin{equation}
\Phi\left(r,t\right)=\frac{\gamma(r)}{\sqrt{8\pi}}e^{i\omega t}\,,\label{ansatz}
\end{equation}
where $\gamma$ is a radial wavefunction and $\omega$ is a real constant representing the angular frequency of the scalar $\Phi$.

%
%
With the metric and scalar field ansätze of Eqs.~\eqref{metric2}-\eqref{ansatz}, the system of Eqs.\eqref{field} to \eqref{kleingordon} provides a system of three independent coupled ODEs for the functions $\sigma, N, \gamma$, 
\beq
N'&=&\frac{1}{r}-N\left(\frac{1}{r}+r\gamma'^2\right)-\frac{\omega^2r\gamma^2}{ N\sigma^2}-\mu^2r\gamma^2\,,\nonumber\\
\sigma'&=&r\frac{\omega^2\gamma^2+N^2\sigma^2\gamma'^2}{N^2\sigma}\,,\nonumber\\ 
\gamma''&=&\frac{\gamma (\mu^2 rN\sigma^2-\omega^2 r)-N\sigma \gamma'\left(r\sigma N'+N(2\sigma+r\sigma')\right)}{rN^2\sigma^2}\,,\nonumber 
\eeq
%
%
%
%
where a prime $'$ denotes a derivative with respect to $r$. The system above has a singular point at the origin $r=0$. To preserve the regularity of the system at the origin, a series expansion of the functions $N$, $\sigma$ and $\gamma$ reveals that these functions must behave as
\beq
m(r)&=&\mathcal O\left(r^3\right)\,,\nonumber \\
\sigma&=&\sigma_0+\mathcal O\left(r^2\right)\,,\label{bcs}\\
\gamma&=&\gamma_0+\mathcal O\left(r^2\right)\,,\nonumber
\eeq
where $\sigma_0$ and $\gamma_0$ are constants. On the other hand, as one seeks a localized solution preserving asymptotic flatness, we require the radial wavefunction $\gamma$ to vanish and the metric functions $(N,\,\sigma) \to 1$ as $r\to\infty$. 

Finding solutions for spherically symmetric and static boson stars consists of solving the above dimensionless equations subjected to the boundary conditions in Eq.~\eqref{bcs}. Due to the complexity of this system, analytical solutions are unattainable, and one usually recurs to numerical methods e.g. shooting methods for the parameter $\omega$ considering a fixed combination of the remaining parameters to find numerical solutions~\cite{Brito:2013xaa,Macedo:2013qea,Macedo:2013jja,Brito:2015yfh}. 

Finally, we note that boson stars have a maximum mass set by the parameter $\mu$:
\be
\frac{M_{\rm max}}{M_\odot}=8\times 10^{-11}\, \left(\frac{\rm eV}{m_s c^2}\right)=8\times 10^{-11}\, \left(\frac{\rm eV}{\hbar \mu c^2}\right)\,.\label{max_mass}
\ee
%
\subsection{The Einstein-Proca theory}\label{sec:eproca}
\subsubsection{Action and equations of motion}

Consider now a theory describing a minimally coupled massive and complex vector field $A^a$. This ``Einstein-Proca'' theory is described by an action of the form 
\begin{equation}\label{action2}
S=\int_\Omega\sqrt{-g}\left(\frac{R}{16\pi}-\frac{1}{4}F_{ab}\bar F^{ab}-\frac{1}{2}\mu^2A_a \bar A^a\right)\,.
\end{equation}
Again, $\mu$ is a mass parameter for the Proca field $A^a$, an overbar ($\bar {\ }$) denotes complex conjugation, and $F_{ab}$ is the electromagnetic tensor defined in terms of $A^a$ as
\begin{equation}\label{defFab}
F_{ab}=\partial_aA_b-\partial_bA_a.
\end{equation}
The theory is controlled by the same dimensionless coupling \eqref{dimensionless_massparameter} as boson stars.

The equations of motion for this theory can be obtained via variations of Eq.\eqref{action2} with respect to the metric $g_{ab}$ and the Proca field $A^a$. The field and Proca equations take thus the respective forms
\beq
R_{ab}-\frac{1}{2}g_{ab}R&=&8\pi T_{ab},\label{fieldeq}\\
T_{ab}&=&-F_{c(a}\bar F^c_{\ b)}-\frac{1}{4}g_{ab}F_{cd}\bar F^{cd} \nonumber \\
&+&\mu^2\left[A_{(a}\bar A_ {b)}-\frac{1}{2}g_{ab}A_c\bar A^c\right],\label{defTab}\\
\nabla_bF^{ab}&=&\mu^2A^a,\label{procaeq}
\eeq
where $T_{ab}$ is the stress-energy tensor for the Proca field and the parenthesis in the indexes denote index symmetrization.

\subsubsection{Equilibrium configurations: Proca stars}

Let us focus in static and spherically symmetric equilibrium solutions of the Einstein-Proca system of Eqs.~\eqref{fieldeq} and \eqref{procaeq}, with the spherically symmetric ansatz of Eq.~(\ref{metric2}). 
Furthermore, we are interested in preserving the time independence of $g_{ab}$ and $T_{ab}$, which can be attained vie the use of the global $U\left(1\right)$ invariance of the action in Eq.\eqref{action2}. To do so, we consider the standing-wave ansatz for the Proca field
\begin{equation}\label{proca}
A_a=e^{-i\omega t}\left(f\left(r\right),i g\left(r\right),0,0\right),
\end{equation}
where $\omega$ is the angular frequency of the Proca field and $f\left(r\right)$ and $g\left(r\right)$ are well-behaved and real functions of the radial coordinate $r$. Inserting Eqs.\eqref{metric2} and \eqref{proca} into the system of Eqs.\eqref{fieldeq} to \eqref{procaeq} one verifies that there are two independent field equations, namely
\begin{equation}\label{field1}
m'=4\pi r^2\left[\frac{\left(f'-\omega g\right)^2}{2\sigma^2}+\frac{1}{2}\mu^2\left(g^2N+\frac{f^2}{N\sigma^2}\right)\right],
\end{equation}
\begin{equation}\label{field2}
\sigma'=4\pi r\mu^2\sigma\left(g^2+\frac{f^2}{N\sigma^2}\right),
\end{equation}
the first of which obtained from the $(t,t)$ component and the second via a combination of the $(t,t)$ and $(r,r)$ components of Eq.\eqref{fieldeq}, as well as two independent Proca equations, which are
\begin{equation}\label{proca1}
\left[\frac{r^2\left(f'-\omega g\right)}{\sigma}\right]'=\frac{\mu^2 r^2 f}{\sigma N},
\end{equation}
\begin{equation}\label{proca2}
\omega g-f'=\frac{\mu^2\sigma^2 Ng}{\omega}.
\end{equation}
These equations form a system of four coupled ODEs for the functions $\sigma$, $N$, $f$ and $g$ which is again singular at the origin. To preserve the regularity of these solutions at the origin, one performs a series expansion of the functions $f\left(r\right)$, $g\left(r\right)$, $m\left(r\right)$ and $\sigma\left(r\right)$ around $r=0$ and obtains the following boundary conditions necessary for a non-singular behaviour
\begin{eqnarray}
f\left(r\right)&=&f_0+\mathcal O\left(r^2\right),\nonumber\\
g\left(r\right)&=&\mathcal O\left(r\right),\label{boundary}\\
m\left(r\right)&=&\mathcal O\left(r^3\right),\nonumber\\
\sigma\left(r\right)&=&\sigma_0+\mathcal O\left(r^2\right),\nonumber
\end{eqnarray}
where $f_0$ and $\sigma_0$ are constants. Furthermore, as we are interested in localized solutions, we want to preserve asymptotic flatness. Thus, we require that the functions $N$ and $\sigma$ approach unity and $f$ and $g$ to vanish as $r\to\infty$. 

Given the complexity of the system of Eqs.~\eqref{field1}-\eqref{proca2}, analytical solutions are unattainable. We thus recur to a numerical integration of the equations subjected to the boundary conditions in Eq.\eqref{boundary} using shooting methods for the parameter $\omega$ with a fixed combination of the remaining parameters $\mu$, $f_0$ and $\sigma_0$. In particular, the parameter $\mu$ can be normalized to $1$ via a redefinition of the radial coordinate and Proca functions in the form
\begin{equation}\label{redefinitions}
x=\mu r, \quad f\left(x\right)=\sqrt{4\pi}f\left(r\right), \quad g\left(x\right)=\sqrt{4\pi}g\left(r\right).
\end{equation}

As with boson stars, Proca stars also have a maximum mass slightly larger than that in Eq.~\eqref{max_mass}~\cite{Brito:2015pxa}.
\subsection{Solutions and fits}\label{sec:solsfits}
\begin{figure}[h]
\includegraphics[scale=0.9]{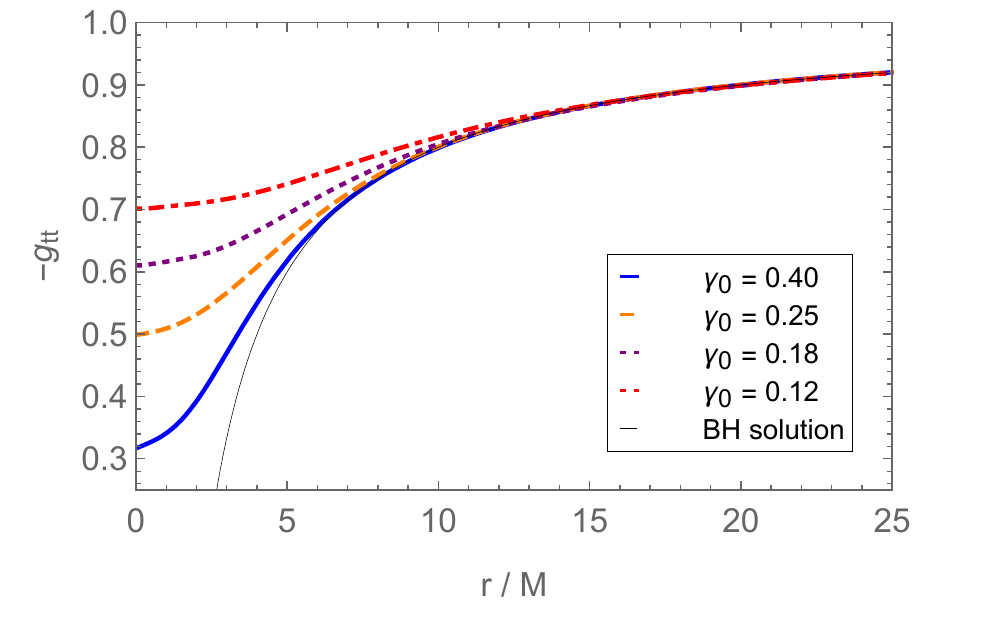}
\includegraphics[scale=0.9]{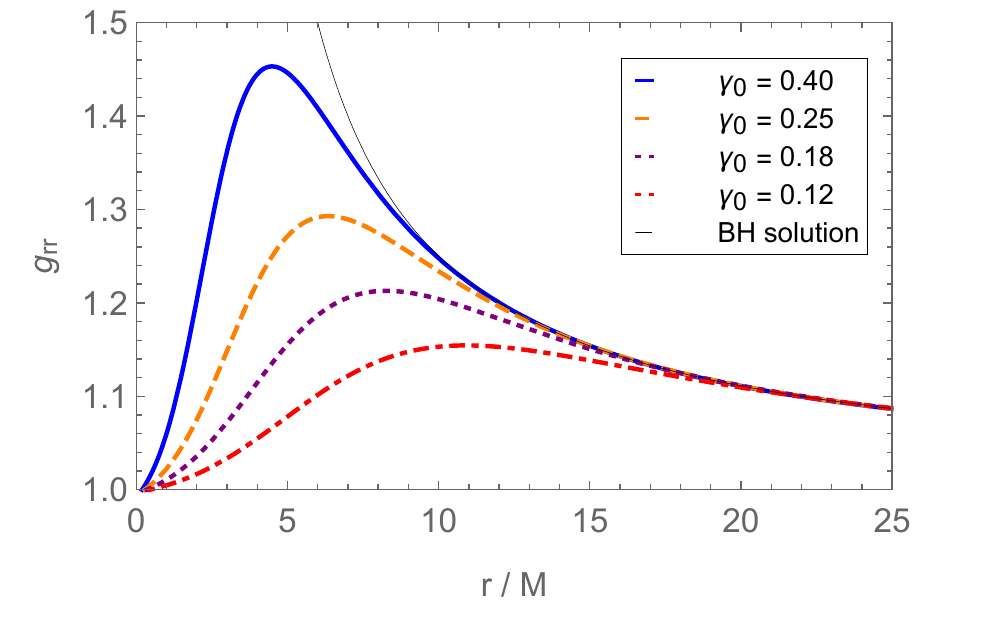}
\caption{Boson star solutions. {\bf Top Panel:} Metric function $g_{tt}$ from Eq.~\eqref{metric2} as a function of the normalized radial coordinate $r/M$. {\bf Bottom Panel:} Metric function $g_{rr}$ from Eq.~\eqref{metric2} as a function of the normalized radial coordinate $r/M$. The thin black line represents the Schwarzschild solution, i.e., $g_{tt}=g_{rr}^{-1}=1-2M/r$. 
Finiteness and positiveness of the metric functions guarantees the non-existence of horizons nor singularities in spacetime.}
\label{fig:BS_sols}
\end{figure}

\begin{figure}[h]
\includegraphics[scale=0.9]{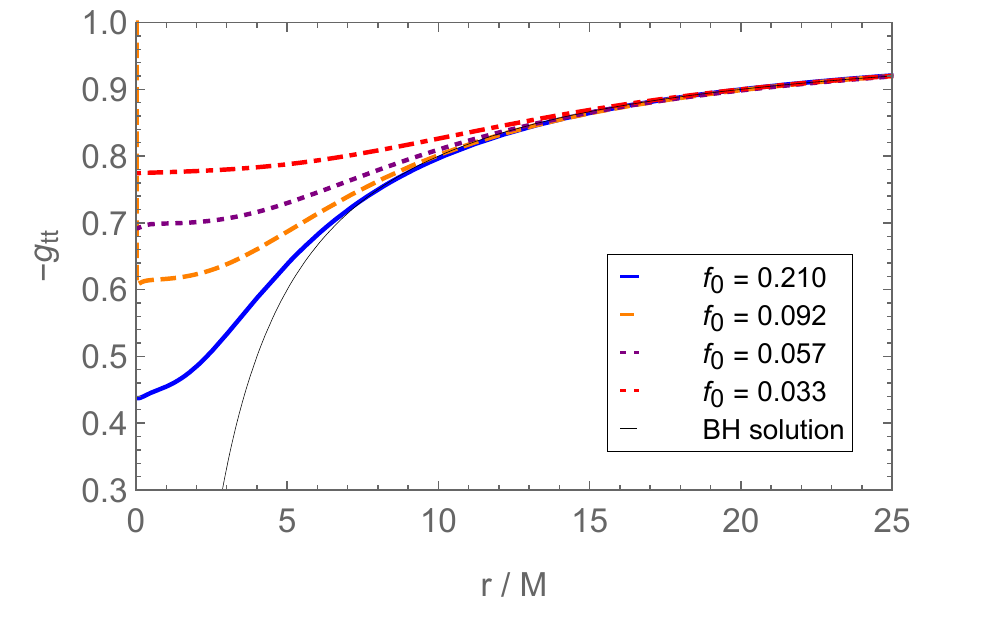}
\includegraphics[scale=0.9]{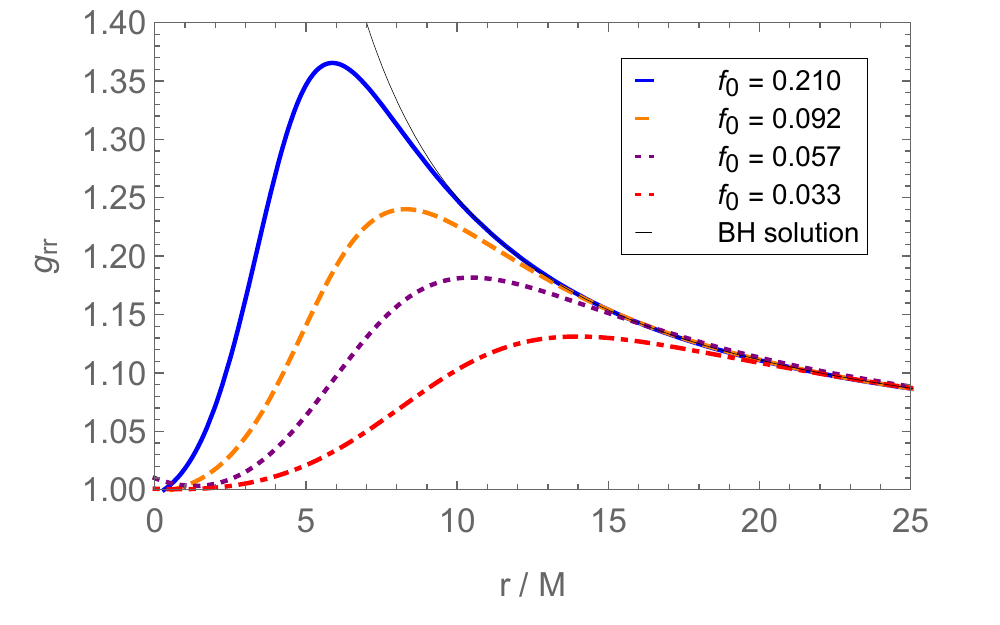}
\caption{Proca stars. {\bf Top Panel:} Metric function $g_{tt}$ from Eq.\eqref{metric2} as a function of the normalized radial coordinate $r/M$. {\bf Bottom Panel:} Metric function $g_{rr}$ from Eq.\eqref{metric2} as a function of the normalized radial coordinate $r/M$. 
The thin black line represents the Schwarzschild solution, i.e., $g_{tt}=g_{rr}^{-1}=1-2M/r$.
Finiteness and positiveness of the metric functions guarantees the non-existence of horizons nor singularities in spacetime.}
\label{fig:metric}
\end{figure}

\begin{table}[ht!]
\begin{tabular}{c c c c c c}
Configuration &$\gamma_0$ & $\mu M$ & $\mu R$ & $R/M$ & $\omega$   \\ \hline
BSC4          &$0.40$ & $0.609$ & $5.46$ & $8.97$ & $0.811$   \\
BSC3          &$0.25$ & $0.632$ & $7.46$ & $11.8$ & $0.864$   \\
BSC2          &$0.18$ & $0.612$ & $9.16$ & $15.0$ & $0.896$   \\
BSC1          &$0.12$ & $0.572$ & $11.1$ & $19.4$ & $0.922$
\end{tabular}
\caption{Relevant parameters describing the boson star configurations considered in this work. Here, $\gamma_0$ is the value of the scalar field at the origin (cf. Eq.~\eqref{ansatz}), the boson star mass is $M$ and it radius $R$ is defined as the radius enclosing $98\%$ of the mass. We will refer to the corresponding configuration acronym in throughout this work.}
\label{tab:bsparam}
\end{table}
\begin{table}[ht!]
\begin{tabular}{c c c c c c}
Configuration &$f_0$ & $\mu M$ & $\mu R$ & $R/M$ & $\omega$ \\ \hline
PSC4          &$0.210$ & $1.04$ & $9.35$ & $8.99$ & $1.28$ \\
PSC3          &$0.092$ & $1.05$ & $12.7$ & $12.1$ & $1.14$ \\
PSC2          &$0.057$ & $1.00$ & $15.1$ & $15.0$ & $1.10$ \\
PSC1          &$0.033$ & $0.925$ & $18.4$ & $19.9$ & $1.06$ \\
\end{tabular}
\caption{Relevant parameters describing the Proca star configurations considered in this work. Here, $f_0$ is the value of $A_t$ component of the vector field at the origin (cf. Eq.~\eqref{proca}), the Proca star mass is $M$ and its radius $R$ is defined as the radius enclosing $98\%$ of the mass.}
\label{tab:proca}
\end{table}
For concreteness, in the remainder of this work, we focus on specific solutions. Namely, we consider four different boson star configurations, detailed in Table \ref{tab:bsparam} and four different Proca star configurations, detailed in Table \ref{tab:proca}. We also list the corresponding configuration acronym which we use throughout this work. The geometry associated with these boson and Proca star configurations are shown in comparison with the Schwarzschild metric in Figs.~\ref{fig:BS_sols} and \ref{fig:metric}, respectively. For completeness, information regarding the scalar and vector field distributions can be found in Appendix \ref{app:fields}. These solutions range from the near-maximum compactness solutions with $R\sim 9M$ to more dilute configurations with $R\sim 20M$. Note that the solutions are indeed asymptotically flat, as the scalar field decays exponentially at large distances.
These solutions have been discussed at length elsewhere~\cite{Liebling:2012fv,Macedo:2013qea,Macedo:2013jja,Brito:2015pxa,Annulli:2020lyc}, we will not dwell on aspects of their structure any further.


The numerical solutions shown in Figs.~\ref{fig:BS_sols} to \ref{fig:metric} are all well-described by analytical expressions of the form
\begin{equation}\label{grraprox}
g_{rr}=\exp\left\{a_7\left[\exp\left(-\frac{1+a_1x+a_2x^2}{a_3+a_4x+a_5x^2+a_6x^3}\right)-1\right]\right\},
\end{equation}
\begin{equation}\label{gttaprox}
g_{tt}=\exp\left\{b_7\left[\exp\left(-\frac{1+b_1x+b_2x^2}{b_3+b_4x+b_5x^2+b_6x^3}\right)-1\right]\right\},
\end{equation}
where $x=\mu r$ is the rescaled radial coordinate, $a_i$ and $b_i$ are constant parameters to be adjusted according to the boundary condition $\gamma_0$ for boson stars and $f_0$ for Proca stars. The values of the parameters $a_i$ and $b_i$ for all of the boson and Proca star configurations considered are summarised in Tables \ref{tab:fitsbs} and \ref{tab:fitsps} in Appendix \ref{app:parameters}, respectively.

Using these parameters, the metric components $g_{rr}$ and $g_{tt}$ of the solutions considered can be approximated by the analytical functions in Eqs.\eqref{grraprox} and \eqref{gttaprox} with relative errors always smaller than $1\%$ and average relative errors in the interval $0<r<50M$ of the order of $0.1\%$. The usefulness of these analytical descriptions of the numerical solutions considered will become evident later on in this work, when we recur to the ray-tracing software GYOTO to generate the observational predictions of isotropically emitting objects orbiting central bosonic stars~\cite{Vincent:2011wz,Vincent:2012kn,Grould:2016emo}. Indeed, some of the necessary inputs to run the code are analytical descriptions of the metric components of the background spacetime, the corresponding Christoffel symbols, and the equatorial orbital velocities as a function of the orbital radius. More information regarding the orbital velocities and the corresponding orbital periods can be found in Appendix \ref{app:periods}.

\section{Orbital motion around bosonic stars}\label{sec:lensing}

Equipped with expressions for the metric and orbital velocity, one can ray-trace orbits of an hot spot around the compact object, using the open source\footnote{Freely available at \href{http://gyoto.obspm.fr}{gyoto.obspm.fr}} GYOTO code~\cite{Vincent:2011wz,Vincent:2012kn,Grould:2016emo}. In this section we present the general behaviour of the ray-traced images, using the Schwarzschild solution as benchmark. We do not include spin in our study because the currently available data is not capable of constraining the BH spin \cite{Gravity2018,Baubock:2020dgq}.

The hot spot is modelled in GYOTO as an isotropically emitting sphere orbiting the central massive object (a boson or Proca star, or a BH) at some constant orbital radius. This model mimics a hot spot in the optically thin accretion disk surrounding the central compact object. We have set the radius of the hot spot to be of $0.5 M$~\citep[in agreement with the upper limit of $0.3\,M$ derived by][]{gillessen06}.
This value is also chosen to be consistent with the literature~\cite{Baubock:2020dgq, Gravity2018}.
The hot spot orbits the central object with an equatorial circular orbit of radius $r$ as described in Appendix~\ref{app:periods}. The angular velocity is computed directly from the metric. 
%
%
The output of GYOTO is a 2D image (with specific intensities $I^\nu_{lm}$) at a given time ($t_k$) of the lensed hot spot, with each of the pixels (i.e., the values of $I^\nu_{lm}$ for some specific $l$ and $m$) representing the specific intensity. These are then converted to $I_{klm}=\Delta\nu I^\nu_{klm} $ cubes, to generate the following observables:
\begin{enumerate}
	\item[a)] time integrated images: 
		\begin{equation}\label{eq:I_lm}
            \langle I\rangle_{lm} = \sum_k  I_{klm};
        \end{equation}
	\item[b)] total temporal fluxes:
	    \begin{equation}\label{eq:F_k}
	        F_k =  \sum_l \sum_m \Delta \Omega I_{klm};
	    \end{equation}
	\item[c)] temporal magnitudes:
	    \begin{equation}\label{eq:m_k}
	        m_k =  -2.5 \log\left(\dfrac{F_k}{\min(F_k)}\right);
	    \end{equation}
	\item[d)] temporal centroids:
    	\begin{equation}\label{eq:c_k}
        	\vec{c}_k=F_k^{-1} \sum_l \sum_m \Delta \Omega I_{klm} \vec{r}_{lm};
    	\end{equation}
\end{enumerate}

\noindent where $\vec{r}_{lm}$ is the position with respect to the centre of the image, $\Delta \nu$ is the spectral width and $\Delta \Omega$ the pixel solid angle (note that the observables in Eqs.~ (\ref{eq:m_k}-\ref{eq:c_k}) are relative and  independent of $\Delta \nu$ and $\Delta \Omega$).

\subsection{Lensed images}
\begin{figure*}[ht!]
    \includegraphics[width=2.0\columnwidth]{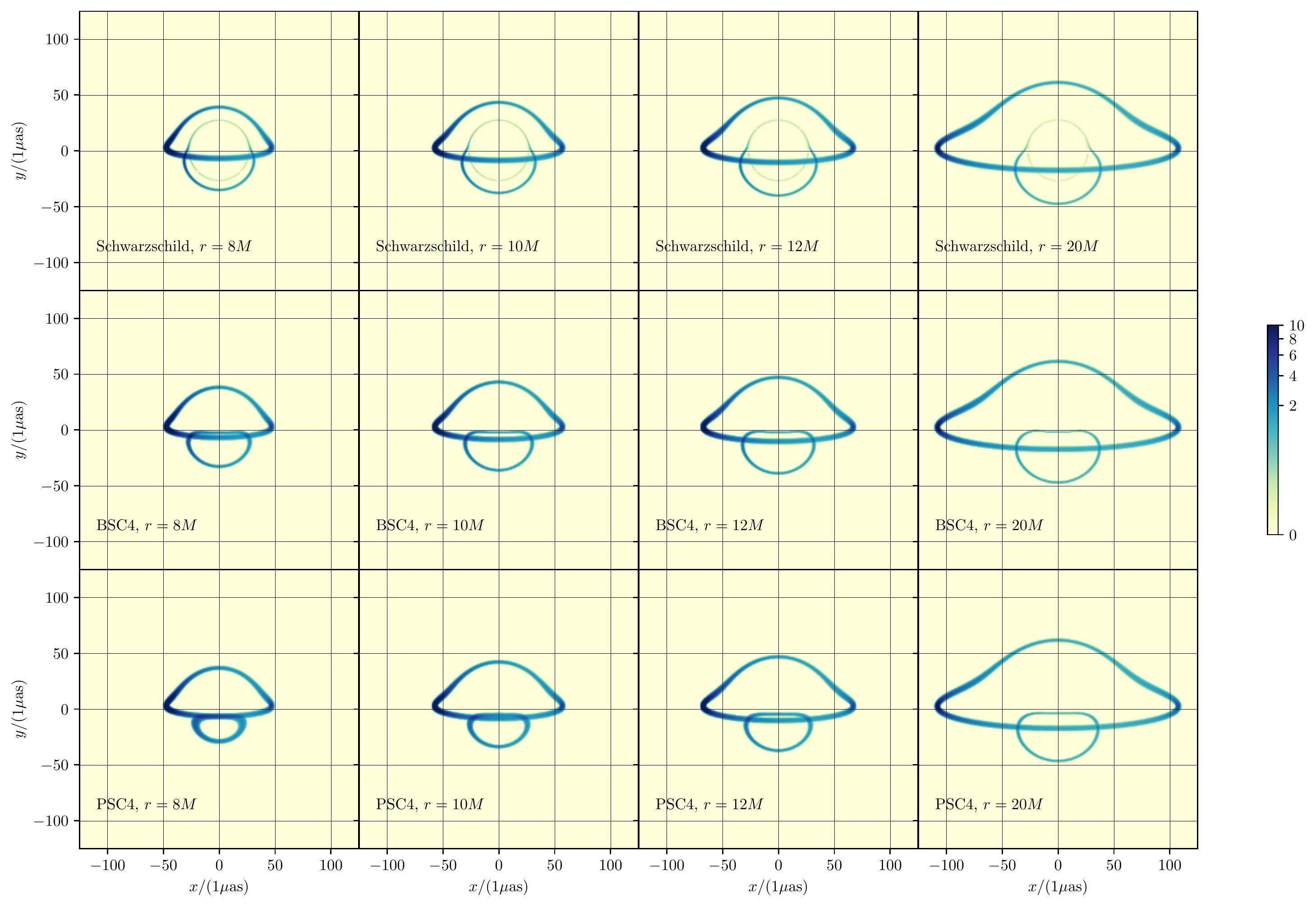}
	\caption{\label{fig:integrated_images_80} Time integrated images 
	for a full orbit, assuming the source to be a spherical luminous spot. The observer inclination is $i=80^\circ$. Different rows show images for different central compact object (Schwarzschild BH, boson star and Proca star, from top to bottom). The columns portray different orbital radius ($8M, 10 M, 12 M, 20M$). See text for discussion.}
\end{figure*}
Figure~\ref{fig:integrated_images_80} shows the time integrated images for the three different central objects we consider (a Schwarzschild BH, a boson star and a Proca star). The images span representative star compactnesses and orbital radii. The observer inclination with respect to the orbital angular momentum is $80^\circ$.

The first row of images in Fig.~\ref{fig:integrated_images_80} depict the gravitational lensing by a Schwarzschild BH, for four different orbital radii of the hot spot.
This image shows well-known features: a) a primary (top) lensed track, corresponding to light traveling  from source to observer without crossing the equatorial plane after emission, thus with a  gravitational deflection which is very small for radiation emitted in the forefront of the BH, and limited for radiation emitted behind the BH as seen from the observer; (b) a secondary (bottom) lensed track, corresponding to light that  makes one half turn around the BH before reaching the observer thus crossing the equatorial plane once after emission; c) beaming emission from the approaching (left) orbit section; d) a faint light ring (so-called "photon ring"), which is barely visible. For sources at larger orbital radii, emitting at the forefront of the BH, the gravitational deflection is smaller, and the time-integrated image approaches that in flat, Minkowski spacetime: the projection of a circle on the sky (notice how the main image size is proportional to the orbital radii). 

Signs of a nontrivial lensing are also seen in the behavior of the secondary image: its size does not increase proportionally to the orbital radii, as only photons which are highly deflected can produce this track, thus always requiring close approach to the central object.

The lensing by boson stars (in particular configuration BSC4, cf. Table~\ref{tab:bsparam}) is summarized in the second row of Fig.~\ref{fig:integrated_images_80}. With the exception of the absence of the light ring (as expected, since only solitonic-type boson stars have light rings~\cite{Macedo:2013jja,Cardoso:2021ehg}), the lensed images are very similar to the Schwarzschild metric. This could be anticipated given how close the metric components are to that of a BH at orbital radii $r\gtrsim 7M$ (cf. Fig.~\ref{fig:metric}; they differ by less than 15\% for orbital radii $r\gtrsim 8M$).

\begin{figure}[ht!]
\begin{center}
\includegraphics[width=\columnwidth]{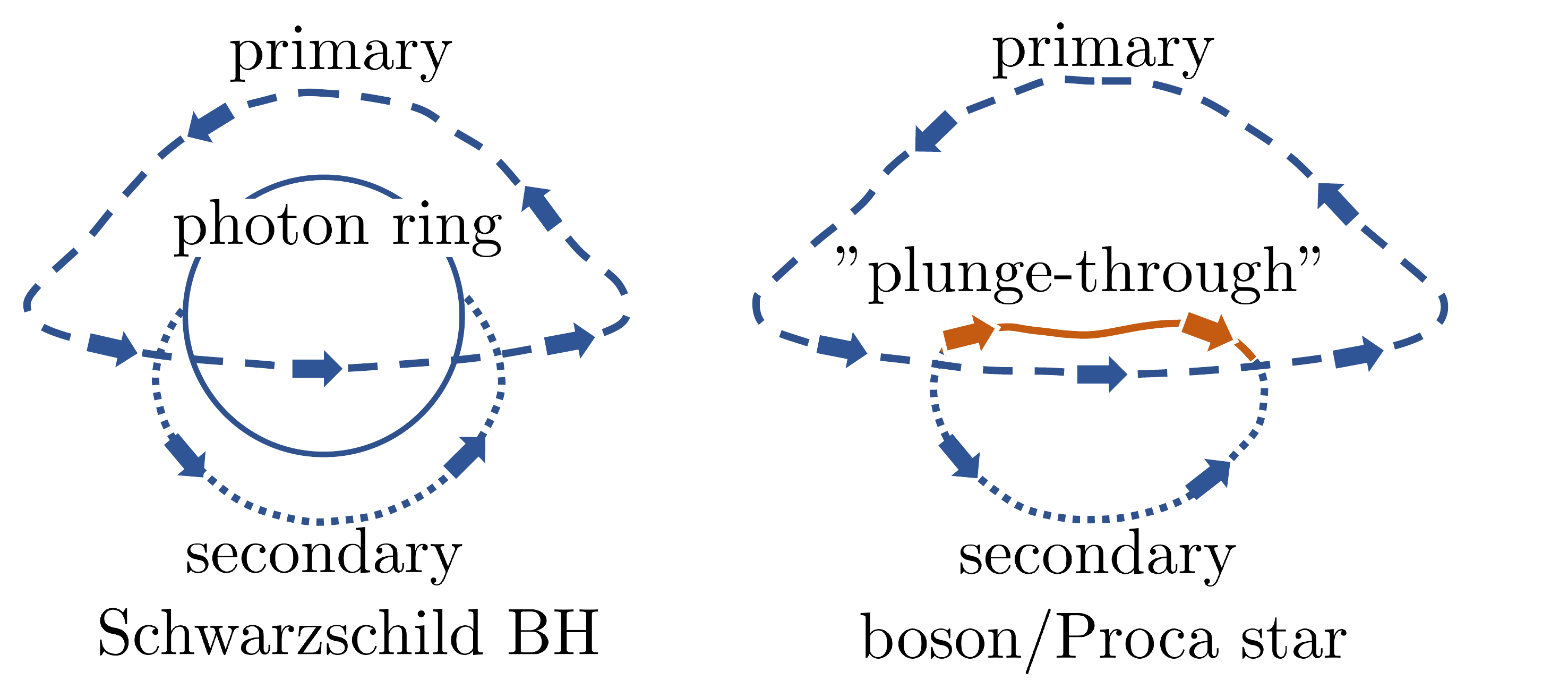}
\end{center}
	\caption{\label{fig:drawing} Drawing of the time integrated images depicting the primary and secondary paths for the compact objects as well as the photon ring in the Schwarzschild BH. The primary track is presented as a dashed line, the secondary is a dotted line, the "plunge-through" and photon ring are in full.}
\end{figure}
A novel aspect of horizonless compact objects is that the secondary track abruptly terminates approximately when it overlaps with the primary track. It then connects to an approximately horizontal track around the center of the image (see
Fig.~\ref{fig:drawing}). This horizontal track is not part of the secondary track, it is due to weakly-lensed photons that go through the compact object (and that obviously do not exist in Shwarzschild because of the presence of the event horizon). We call this
feature the "plunge-through" image
(see Fig.~\ref{fig:drawing}). Note that while it appears that the secondary track and the plunge-through track combine into a single continuous track, this is not true in terms of motion. Indeed, while the primary track corresponds to a full orbit of the source around the central bosonic star, meaning that a primary image is always present in the observer's screen independently of the position of the source, the same is not true about the secondary and plunge-through tracks. These two tracks are only visible when the source is moving behind the central bosonic star as seen from the observer's point of view, and they produce two images that move in the same direction, eventually merging together and disappearing. This feature will be clarified in what follows via the analysis of temporal fluxes.


The third, bottom row in Fig.~\ref{fig:integrated_images_80} depicts the images when the central objects is a Proca star (PSC4, cf. Table~\ref{tab:proca}), with a hotspot at similar orbital radii. The images are remarkably similar to the boson star case. There exists a slight difference in the secondary image track lensing. We attribute this difference to the slight difference in the geometries: while both the Proca and boson star spacetimes are similar to their Schwarzschild counterparts, the metric component $g_{rr}$ of boson stars goes to larger values than its Proca counterpart (this is apparent in Figs.~\ref{fig:BS_sols}-\ref{fig:metric}). In other words, the core of boson stars is more compact than that of Proca stars, the latter therefore require closer approach to lense signals back to the observer.
In Appendix~\ref{app:integrated} the integrated images are presented for $i=20^\circ,50^\circ, 90^\circ$.

\begin{figure*}[ht!]
\includegraphics[width=2\columnwidth]{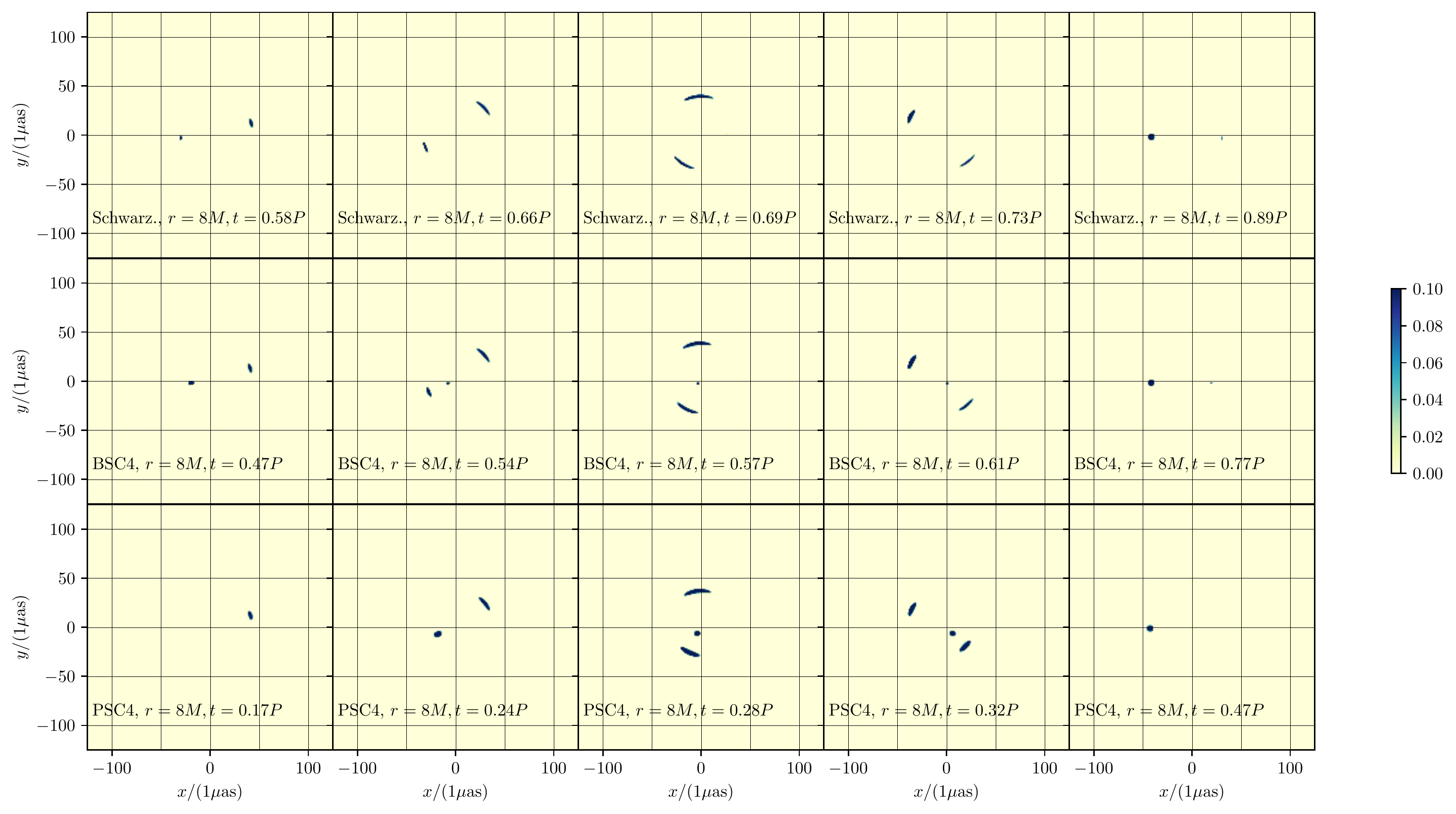}
	\caption{\label{fig:lensing} Lensed images for different times. Top row: Schwarzschild BH, middle row: boson star, bottom row: Proca star. The inclination is $i=80^\circ$, the orbital radius is $r=8M$. See text for details.}
\end{figure*}
We now consider the temporal sequence of lensed images, summarized in Fig.~\ref{fig:lensing}, for the same central objects and inclination angle ($i=80^\circ$). We focus on an orbital radius $r=8M$ and a hot spot orbiting counter-clockwise. The observation times of each frame were chosen such that the primary image is at the same sky position for all geometries. Because the orbital periods are different for each metric, these positions occur at different times. The time frame initiates for all metrics at the same position in the orbit where the hot spot is near $\mathrm{max}(x)$ (the rightmost position in the image).

When the star is at the rightmost position, the images as seen by far-away observers are shown in the first column (left) of Fig.~\ref{fig:lensing}. The observer always sees a primary image of the spot, plus a secondary that is almost always visible, whatever the central compact object. The  primary image relates to photons which travel directly from the hot spot to the observer. The dimmer secondary image corresponds to lensed photons, which were initially travelling to the left, but were deflected by the compact object. Such a secondary image is also present for Proca stars, but due to its core being less compact (see discussion above), it is much fainter.

However, the boson and Proca stars show a unique feature, a new image -- the ``plunge-through'' image -- associated with photons crossing the center of the compact object. We insist that this is not a tertiary image: these photons are actually very weakly lensed as they go "straight" from behind the compact object to the observer, travelling through in the absence of an event horizon. They are weakly lensed because the impact parameter is small. There are other higher-order images for BHs, corresponding to photons circling the light ring a number of times; however, when the central object is horizonless, the ``plunge-through'' image is located within what would be a ``shadow region'' for the BH, allowing light to cross a region which would be located inside the horizon in a BH spacetime. In the following columns as the spherical spot orbits behind the compact object, the lenses follow the usual behaviour, with the central spot (plunge-through image of the boson and Proca stars) moving through the central object and progressively merging into the secondary image. As will be explained later the secondary and "plunge-through" images always appears simultaneously and evolve in parallel.
Moreover, we note that the horizonless secondary image is less extended than its Schwarzschild counterpart: it lacks the top part of it, along the Schwarzschild photon ring. This is because the horizonless spacetimes do not possess any photon orbits, so that there do not exist any extremely bent photons like that forming the top part of the
Schwarzschild secondary image.

\begin{figure}[ht!]
    \centering
    \includegraphics[scale=0.43]{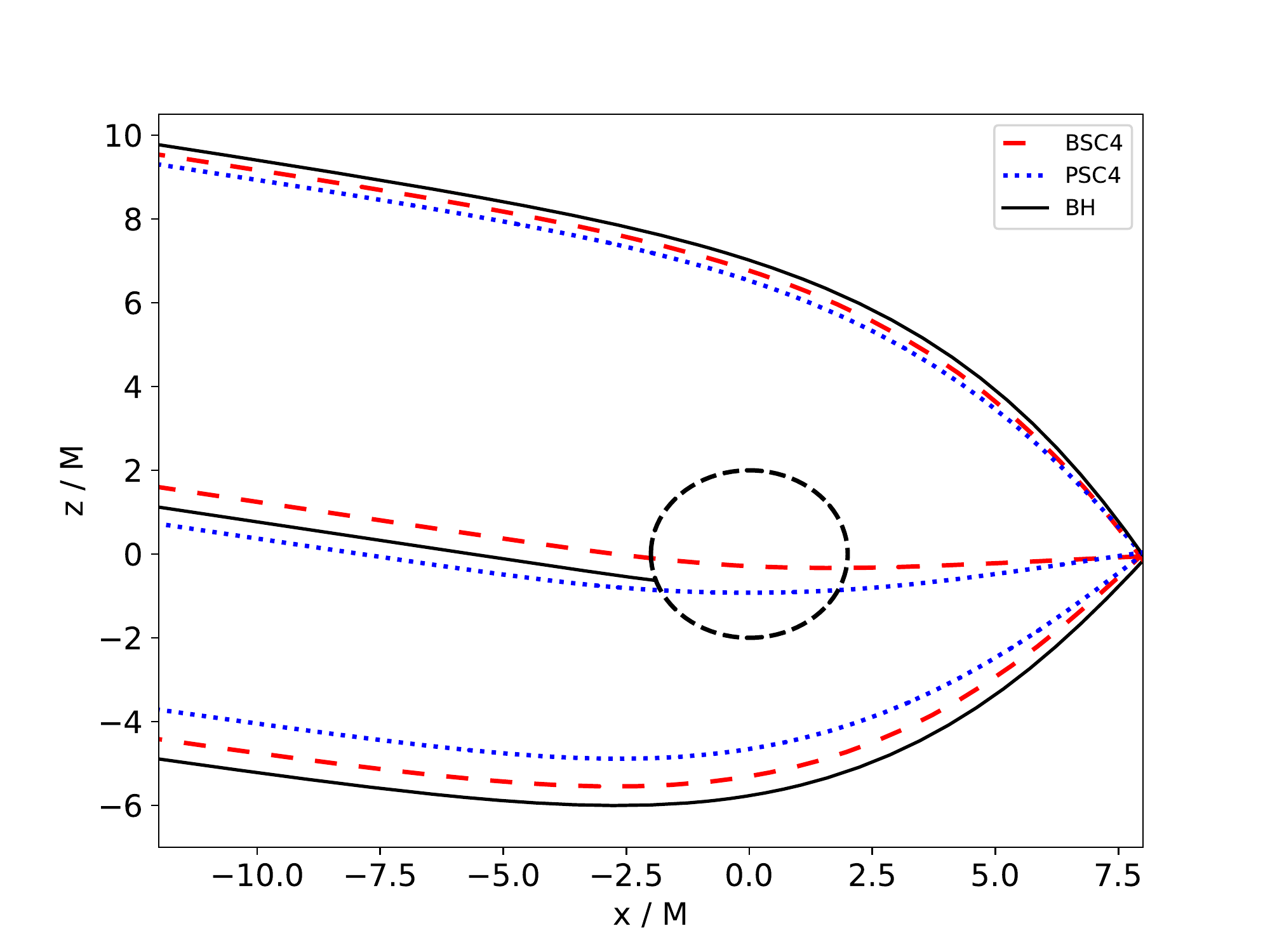}
    \caption{Geodesics connecting the hot spot at $x=8M$ and $z=0$ to the observer at $x=-1000M$ with an observation angle of $i=80^\circ$ for the BSC4, PSC4, and Schwarzschild configurations. The black dashed circle corresponds to the event horizon. The third source does not appear in the lensed images of the Schwarzschild case because the associated geodesic would cross the event horizon.}
    \label{fig:geod1}
\end{figure}
To make the above conclusions more clear, let us consider the structure of geodesics received by the observer. For concreteness, focus on an observation angle of $i=80^\circ$, an orbital radius for the hot spot of $r=8M$, and an appropriate instant of time for which there are the primary and secondary images for all spacetime, plus the "plunge-through" image for the boson and Proca stars spacetimes. In Fig.~\ref{fig:geod1}, we trace a total of nine geodesics in this configuration: three associated to a pixel in the primary images, three associated to a pixel in the secondary images, two associated to pixels in the  "plunge-through" image appearing in the boson and Proca star lensed images, and one for an empty pixel in the Schwarzschild case in the region corresponding to the "plunge-through" image of the horizonless spacetimes. The horizontal axis represents the $x$ coordinate, and the vertical axis represents the $z$ coordinate, both in units of $M$. The observer stands to the left side of the image, at $x=-1000M$, and at this particular instant the hot spot stands at $x=8M$ and $z=0$, i.e., at the equatorial plane where the geodesics converge. One can verify that the geodesics associated to the "plunge-through" image in the bosonic star configurations correspond to geodesics that, in the Schwarzschild case, would cross the horizon, and hence the reason for their absence in BH spacetimes.

\subsection{Temporal fluxes and centroids}
%
\begin{figure*}[h!]
	\includegraphics[width=1.55\columnwidth]{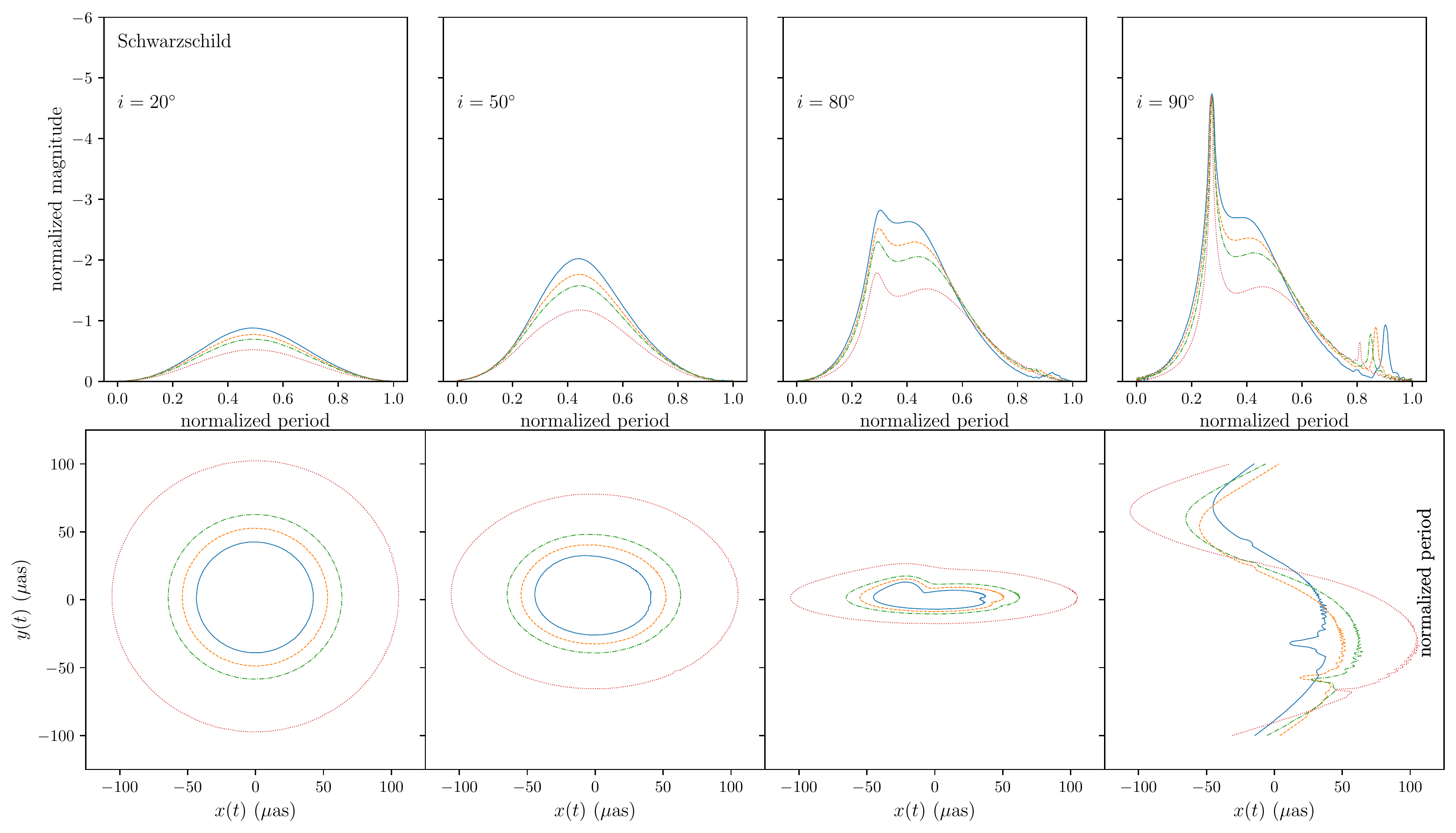}\\
	\includegraphics[width=1.55\columnwidth]{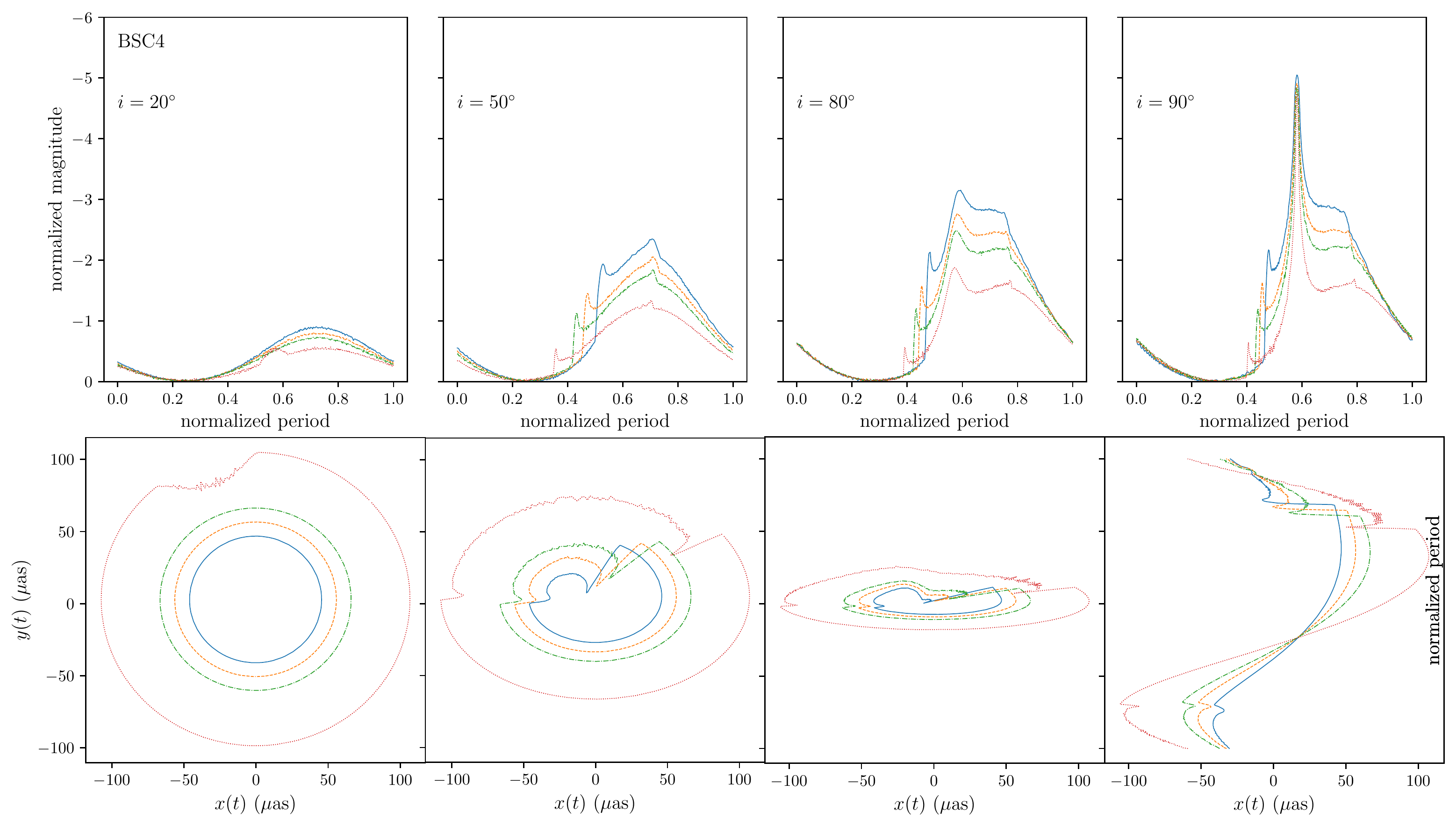}\\
		\includegraphics[width=1.55\columnwidth]{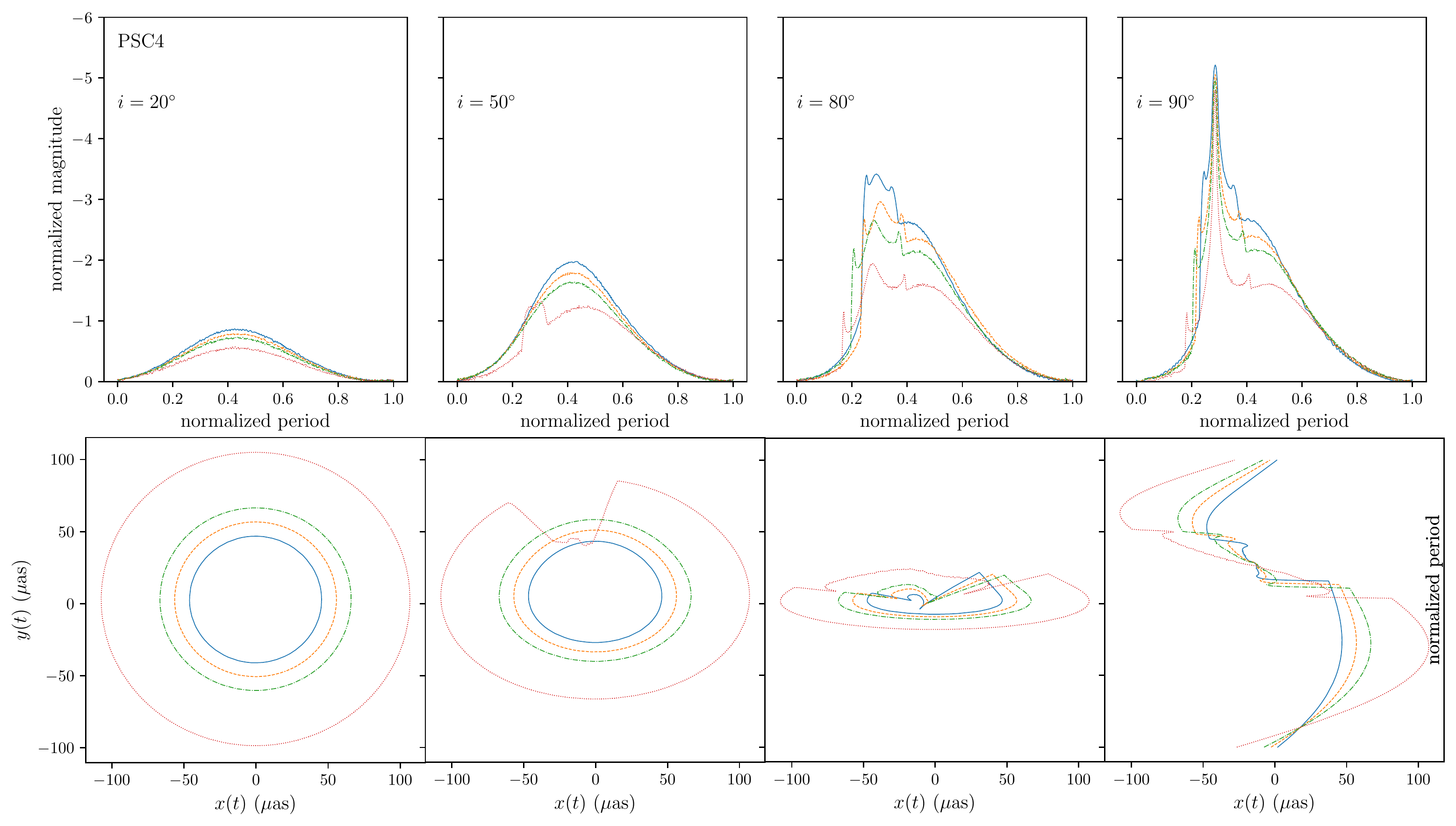}\\
	\caption{\label{fig:centroid_C4} Temporal magnitude $m_k$ and temporal centroid $\vec{c}_k$ for Schwarzschild and most compact boson and Proca stars  (cf. Tables~\ref{tab:bsparam} and \ref{tab:proca}). For the $i=90^\circ$ centroid, the vertical axis is time. See text for details.}
\end{figure*}
The temporal flux $F_k$ in Eq.~(\ref{eq:F_k}) provides complementary information to the one discussed above. Figure~\ref{fig:centroid_C4} shows the temporal magnitude $m_k$ defined in Eq.~(\ref{eq:m_k}) as function of time (normalized by an orbital period), and the temporal centroids $\vec{c}_k$  (defined in Eq.~(\ref{eq:c_k}) for different central objects. Our approach is similar to the one presented in \cite{2009ApJ...692..902H}.
Consider the BH case first, shown in the top panels. They show a ``double hump,'' apparent for higher inclinations but always present. These are caused by the secondary image contributing to the flux. Numerical noise is evident in the low flux region, a fraction at least arises from flux in the light ring pixels. As might be anticipated from the previous discussion, a extra peak arises when the central object is a boson or Proca star, caused by the "plunge-through"  image corresponding to light rays crossing the object. Let's consider the innermost orbit at $i=80^\circ$ for BSC4. By comparing with Fig.~\ref{fig:lensing} it is apparent that at $t=0.47P$ the secondary image appears and a sudden increase in flux takes place. The secondary includes both the classical one as well as the "plunge-through". As time passes the "plunge-through"  detaches and proceeds to the right of the image, it's angular size and brightness decreasing with time, creating the extra peak in the light curve.

With regards to the temporal centroid positions a new signature is present, a shift in the centroid towards the centre for a fraction of the orbit that depends on the inclination and orbital radius. This shift is present for the BS cases at every inclination. However, at $20^\circ$, it is only found for the biggest orbital radius (left panel, red track), while it is present for all orbital radii at higher
inclinations. 
The situation is similar for PS cases, but the centroid shift starts to
be visible only for inclination $\geq 50^\circ$.
This centroid shift is due to the appearance of the secondary image and of
the plunge-through image, that appear in the central part of the image and
thus push the centroid towards the center. Let us explain why this
centroid shift is only present for our horizonless spacetimes.



\begin{figure*}
    \centering
    \includegraphics[scale=0.8]{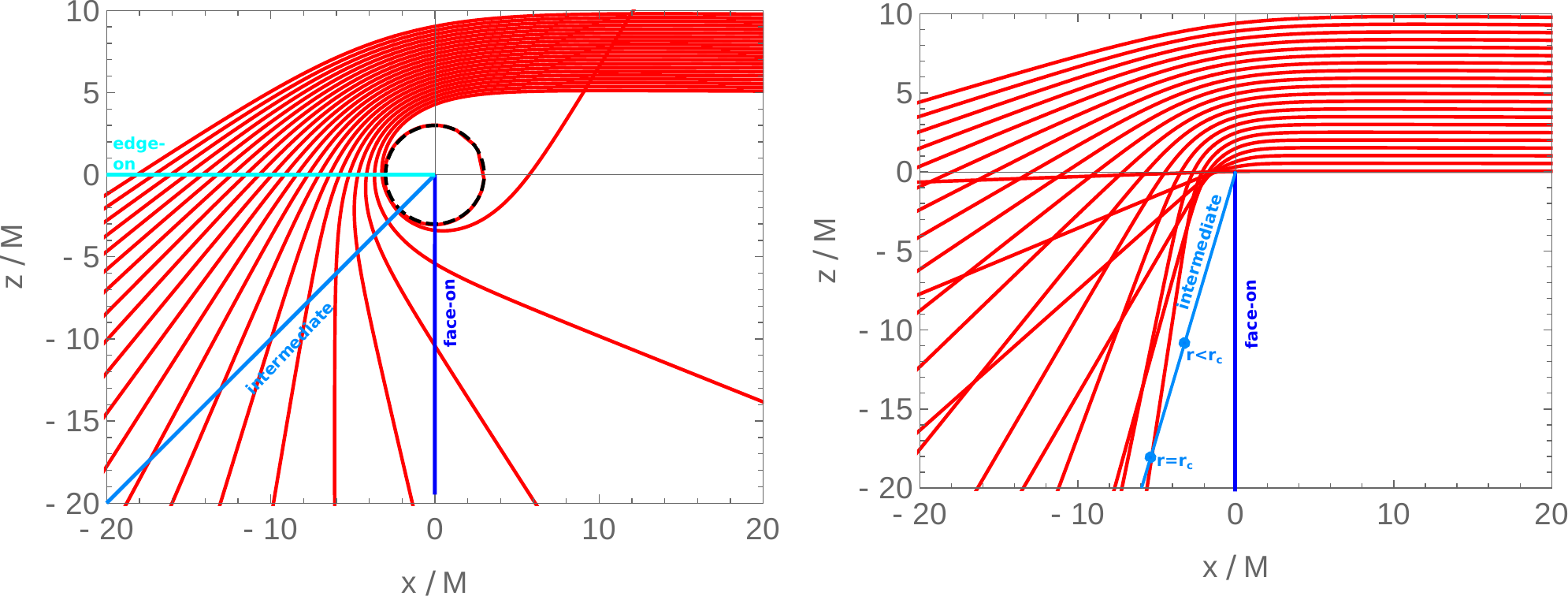}
    \caption{Geodesic congruences in the Schwarzschild spacetime (\textbf{Left panel}) and in the BSC4 boson star spacetime (\textbf{Right panel}). The dashed black circle in the upper panel corresponds to the light-ring. The observer is located towards the right. The position of the equatorial plane is represented in blue with a different hue depending on the inclination of the observer. The entire equatorial plane outside the light-ring is projected onto the observer's screen for the Schwarzschild case independently of the observation angle. For the BSC4 case, only a certain portion
    $r>r_c$ of the equatorial plane is projected onto the observer's screen, where $r_c$ is a critical radius  that varies with the
    inclination angle. 
    In particular, this image shows that for low inclination angles the secondary image is always absent in the BSC4 case.}
    \label{fig:congruences}
\end{figure*}

This discrepancy between the observations for Schwarzschild and bosonic star spacetimes arises from the fact that the bosonic stars studied in this work do not present a light-ring. In the Schwarzschild case, the strong null geodesic curvature in the vicinity of the light-ring leads to the entire equatorial plane outside the light-ring being projected onto the observer's screen as a secondary image, independently of the observation angle (see Fig.~\ref{fig:congruences}, left panel). This is not so in the bosonic star cases where the situation depends a lot on the inclination:
\begin{itemize}
\item for very low inclination, no secondary image of the equatorial plane can reach the observer; 
\item only above a certain critical inclination angle does a
secondary image appear for some part of the equatorial plane. Only a portion of the equatorial plane is projected onto the observer's screen, from some critical radius $r_c$ up to infinity; 
\item only in the limit of edge-on inclination does the full equatorial plane gets projected to a secondary image on
sky whatever the radius.
\end{itemize}
These various situations are illustrated in 
Fig.~\ref{fig:congruences}, right panel.
An analysis of the critical inclination angles for which the secondary image appears can be found in Appendix \ref{app:angles}.
Furthermore, notice that in the case of the BSC4 configuration, for any point in the equatorial plane at a radius larger than the critical radius $r_c$, there are always two geodesics connecting that point to the observer: one associated to the secondary image and one associated
to the plunge-through image. So the secondary and plunge-through images
always come together.
These findings allow to understand the difference between the
horizonless centroid tracks and that of Schwarzschild.
When the inclination becomes high enough that the horizonless
spacetime allows the creation of a secondary image, and only
for radii bigger than $r_c$, then the secondary/plunge-through pair
appears and shifts the centroid towards the center.
In Schwarzschild, the secondary image being always present, there
is not such an effect.


\section{Conclusions}\label{sec:concl}
Our results indicate clear signatures of strong lensing by horizonless objects, in particular a ``plunge-through'' image, corresponding to light crossing the object. In the setup we studied, the matter composing the horizonless, compact object does not couple to light. Thus, our setup could describe objects which are dark matter clumps mimicking BHs. 
As soon as couplings are allowed, the extra image -- corresponding to photons crossing the central massive object -- is either blurred or strongly suppressed depending on the coupling strength. Note however that, due to the uneven scalar/vector field densities throughout the bosonic star configurations considered, one does not expect this effect to affect equally the whole plunge-through track, or to be at all restricted to this track. Depending on the radius and compacticity of the bosonic star considered, one might still have a non-neglectable bosonic field density along the path crossed by the photons that produce the primary and secondary images (this can happen e.g. for the bosonic stars with a larger radius). Furthermore, since the density of the bosonic field peaks at the center and decays exponentially for large radii, one does not expect the couplings to induce an abrupt decrease in the intensity of the images, but instead a fading-out effect from the peripheral region to the center, an effect that is absent in black-hole spacetimes and whose detection would provide a strong indication of the presence of these configurations. For axionic-type couplings for example, other effects could occur including birrefringence~\cite{PhysRevD.41.1231,Harari:1992ea,Chen:2021lvo}, which can lead to frequency-independent oscillations in
the electric vector position angle, and possibly to characteristic signals.
When couplings to baryonic matter are too strong, the "plunge-through" image simply no longer exists.

In a forthcoming paper these results will be put to test in the context of the  detection of orbital motion in SgrA* flares \cite{Gravity2018,Baubock:2020dgq}. Continuous monitoring with the GRAVITY+ instrument will detect a large sample of flare orbits allowing the characterisation of astrophysical effects and unveiling new tests on the nature of the Galactic Centre compact object.

\begin{acknowledgments}
JLR was supported by the European Regional Development Fund and the programme Mobilitas Pluss (MOBJD647).
V. C. is a Villum Investigator supported by VILLUM FONDEN (grant no. 37766) and a DNRF Chair supported by the Danish Research Foundation. V.C. acknowledges financial support provided under the European
Union’s H2020 ERC Advanced Grant “Black holes: gravitational engines of discovery” grant agreement
no. Gravitas–101052587.
This project has received funding from the European Union's Horizon 2020 research and innovation programme under the Marie Sklodowska-Curie grant agreement No 101007855.
We thank FCT for financial support through grants UIDB/00099/2020, PTDC/MAT-APL/30043/2017 and PTDC/FIS-AST/7002/2020.
\end{acknowledgments}

\bibliographystyle{apsrev4-1}
\bibliography{HotSpot}

\appendix

\section{Scalar and vector field distributions}
\label{app:fields}

\begin{figure}
    \centering
    \includegraphics[scale=0.8]{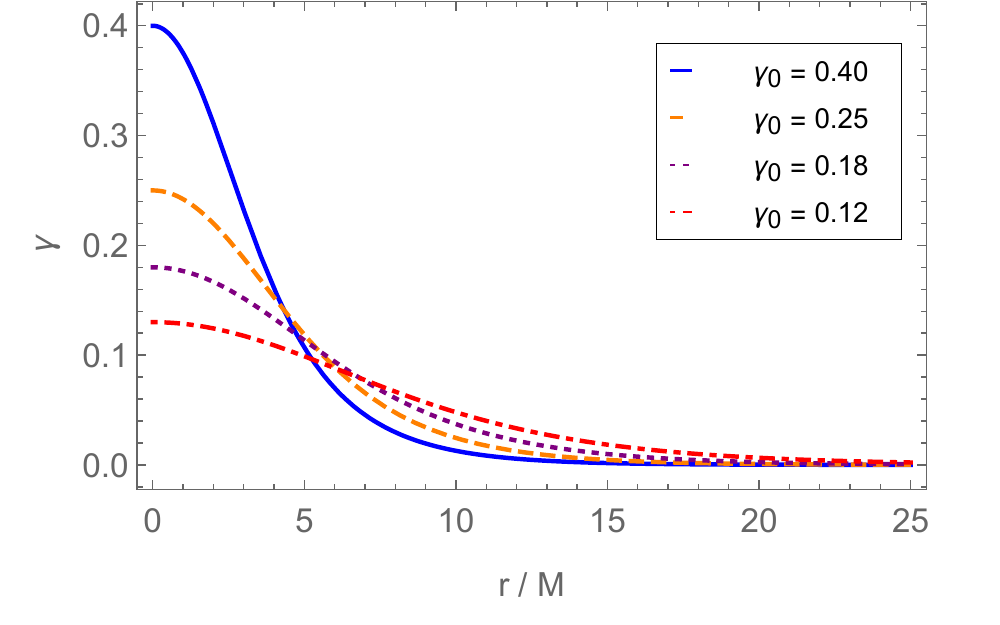}
    \caption{Scalar field $\gamma$ function as a function of the normalized radial coordinate $r/M$. The exponential decay at large radii confirms the confinement of the boson star in a finite region of space near the origin.}
    \label{fig:scalar}
\end{figure}

\begin{figure}[h]
\includegraphics[scale=0.7]{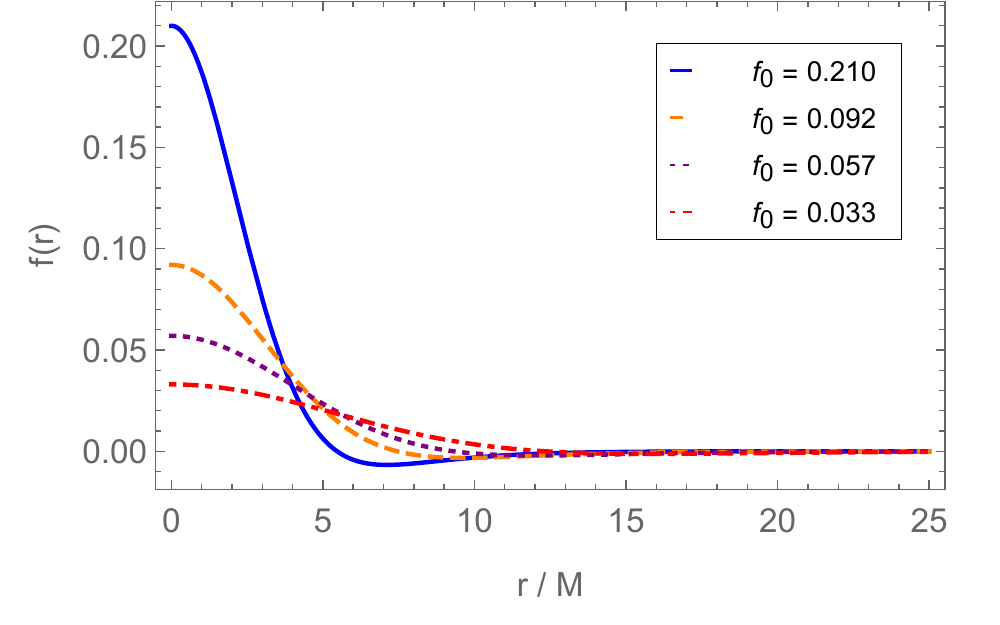}
\includegraphics[scale=0.7]{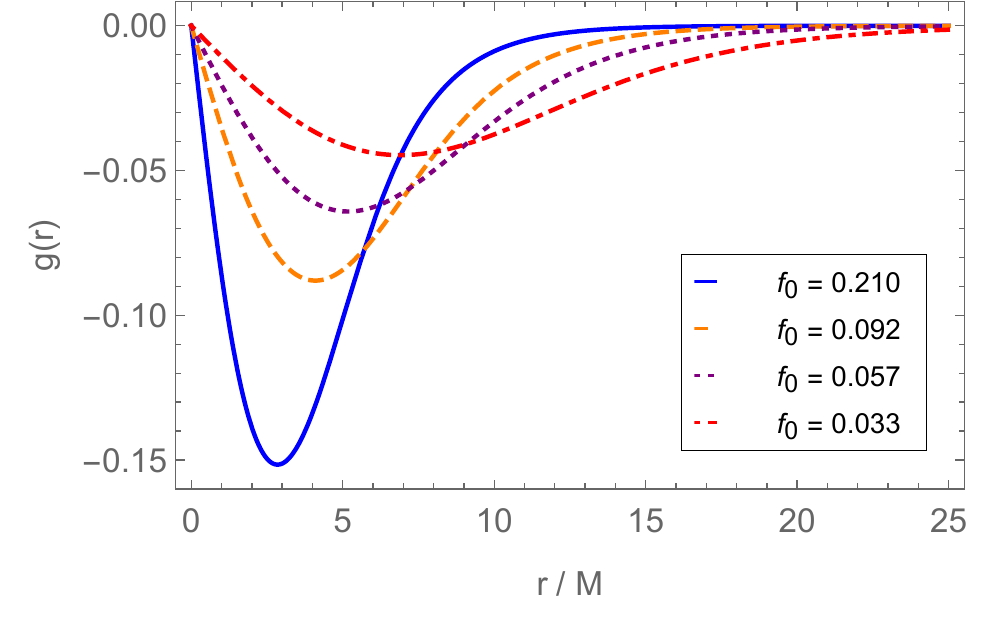}
\caption{{\bf Top Panel:} Vector field $f$ function as a function of the normalized radial coordinate $r/M$. {\bf Bottom Panel:} Vector field $g$ function as a function of the normalized radial coordinate $r/M$.
The exponential decays of both functions at large radii confirm the localization of the Proca star in a finite region of space near the origin.}
\label{fig:vector}
\end{figure}

In Sec.\ref{sec:solsfits} we have introduced four boson star and four Proca star configurations, whose parameters have been detailed in Tables \ref{tab:bsparam} and \ref{tab:proca}, and associated metric components $g_{tt}$ and $g_{rr}$ have been plotted in Figs. \ref{fig:BS_sols} and \ref{fig:metric}, respectively. For completeness, in this section we also provide the plots for the scalar and vector field distributions associated with each of the solutions provided. In Fig.\ref{fig:scalar}, we plot the redefined scalar field $\gamma$, whereas in Fig.\ref{fig:vector} we plot both the functions $f$ and $g$, i.e., the time and radial components of the vector field $A^\mu$, all as functions of the normalised radial coordinate $r/M$. The exponential decay of the solutions at large radii confirms the localization of the solutions in a finite region of space near the origin.

\section{Fit parameters for the solutions considered}
\label{app:parameters}
In this section, we provide the values of the fit parameters $a_i$ and $b_i$ in Eqs.\eqref{grraprox} and \eqref{grraprox} for the boson and Proca star configurations considered. For the boson stars detailed in Table \ref{tab:bsparam}, the associated fit parameters are given in Table \ref{tab:fitsbs}, whereas for the Proca stars detailed in Table \ref{tab:proca}, the associated fit parameters are given in Table \ref{tab:fitsps}.

\begin{table*}[h]
\begin{tabular}{c c c c c c c c}
$\gamma_0$ & $a_1$ & $a_2$ & $a_3$ & $a_4$ & $a_5$ & $a_6$ & $a_7$ \\ \hline 
$0.40$ & $-8.38$ & $-1.77$ & $6.08$ & $-0.204$ & $1.32$ & $0.0750$ & $0.0536$ \\
$0.25$ & $-5.63$ & $-0.797$ & $6.59$ & $-0.295$ & $0.647$ & $0.0244$ & $0.0412$ \\
$0.18$ & $-4.55$ & $-0.457$ & $6.22$ & $-0.163$ & $0.387$ & $0.00991$ & $0.0291$ \\
$0.12$ & $-3.78$ & $-0.261$ & $5.63$ & $-0.0354$ & $0.231$ & $0.00378$ & $0.0192$\\
\end{tabular}
\ \\
\begin{tabular}{c c c c c c c c}
$\gamma_0$ & $b_1$ & $b_2$ & $b_3$ & $b_4$ & $b_5$ & $b_6$ & $b_7$ \\ \hline
$0.40$ & $0.269$ & $0.211$ & $0.304$ & $0.290$ & $0.0250$ & $0.209$ & $1.19$ \\
$0.25$ & $0.107$ & $0.0492$ & $0.702$ & $0.115$ & $0.0729$ & $0.0346$ & $0.916$ \\
$0.18$ & $6.05$ & $1.77$ & $-0.0429$ & $2.15$ & $-0.209$ & $0.764$ & $0.494$ \\
$0.12$ & $3.44$ & $0.844$ & $-0.0264$ & $1.32$ & $-0.0993$ & $0.280$ & $0.355$ \\ 
\end{tabular}
\caption{Values of the parameters $a_i$ and $b_i$ in Eqs.\eqref{grraprox} and \eqref{gttaprox} for the four boson star solutions considered in Table \ref{tab:bsparam}. These combinations allow for the approximation of the metric components $g_{rr}$ and $g_{tt}$ with relative errors smaller than $1\%$ and average relative errors of the order of $0.1\%$ in the range $0<r<50M$.}
\label{tab:fitsbs}
\end{table*}
\begin{table*}[h]
\begin{tabular}{c c c c c c c c}
$f_0$ & $a_1$ & $a_2$ & $a_3$ & $a_4$ & $a_5$ & $a_6$ & $a_7$ \\ \hline 
$0.210$ & $-2.71$ & $-1.40$ & $144$ & $-25.7$ & $2.43$ & $0.413$ & $0.644$ \\
$0.092$ & $-1.09$ & $-0.999$ & $77.7$ & $-10.9$ & $0.935$ & $0.0785$ & $0.176$ \\
$0.057$ & $-1.03$ & $0.384$ & $67.9$ & $-0.660$ & $-0.958$ & $0.138$ & $-0.677$ \\
$0.033$ & $-1.48$ & $1.00$ & $422$ & $-35.3$ & $-0.103$ & $0.199$ & $-0.378$\\
\end{tabular}
\ \\
\begin{tabular}{c c c c c c c c}
$f_0$ & $b_1$ & $b_2$ & $b_3$ & $b_4$ & $b_5$ & $b_6$ & $b_7$ \\ \hline
$0.210$ & $5.75$ & $1.10$ & $-0.0429$ & $2.15$ & $-0.0625$ & $0.484$ & $0.827$ \\
$0.092$ & $23.3$ & $3.06$ & $-0.144$ & $7.19$ & $-0.325$ & $0.827$ & $0.498$ \\
$0.057$ & $15.5$ & $1.87$ & $-0.104$ & $5.23$ & $-0.428$ & $0.391$ & $0.368$ \\
$0.033$ & $-0.0103$ & $0.00461$ & $0.812$ & $0.00440$ & $0.00281$ & $0.000799$ & $0.361$ \\ 
\end{tabular}
\caption{Values of the parameters $a_i$ and $b_i$ in Eqs.\eqref{grraprox} and \eqref{gttaprox} for the four Proca star solutions considered in Table \ref{tab:proca}. These combinations allow for the approximation of the metric components $g_{rr}$ and $g_{tt}$ with relative errors smaller than $1\%$ and average relative errors of the order of $0.1\%$ in the range $0<r<50M$.}
\label{tab:fitsps}
\end{table*}

\section{Orbital velocity and period compared to the BH case}
\label{app:periods}

Let us consider a massive particle undergoing circular orbital motion around a central massive object, in the geodesic approximation (i.e., there is no backreaction in the spacetime). Circular orbits are characterized by the conditions $\dot r=\ddot r =0$, where a dot denotes a derivative with respect to the affine parameter of the geodesics. Furthermore, given the spherical symmetry of the problem, one can always restrict the analysis to the equatorial plane without loss of generality, i.e., by considering $\theta=\pi/2$ and $\dot \theta=\ddot\theta=0$, which can be shown to satisfy the geodesic equation for the angle $\theta$. The angular velocity $\Omega_c$ of an orbit around a central object described by a given metric $g_{ab}$ satisfying these requirements is given by
\begin{equation}\label{orbvel}
\Omega_c = \frac{d\phi}{dt}= \sqrt{\frac{1}{2r}\frac{d}{dr}\left(g_{rr}\right)}.
\end{equation}
The orbital period can then be computed via $T=2\pi/\Omega$. The angular velocities for each of the cases considered, i.e. $\Omega_{\text{BS}}$ for the boson star, and $\Omega_{\text{PS}}$ for the Proca star, are given in terms of their respective metric functions as
\begin{equation}
\Omega_{\text{BS}}=\sqrt{\frac{e^{\nu\left(r\right)}\nu'\left(r\right)}{2r}},
\end{equation}
\begin{equation}
\Omega_{\text{PS}}=\sqrt{\frac{\sigma}{2r}\left(2\sigma'N+\sigma N'\right)}.
\end{equation}

\begin{figure}
\includegraphics[scale=0.9]{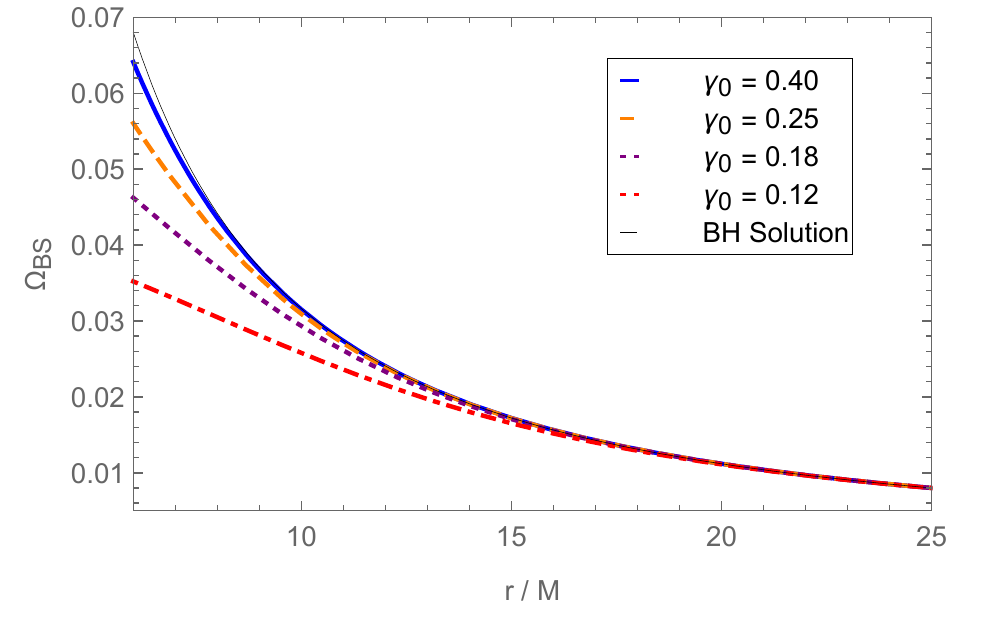}
\includegraphics[scale=0.9]{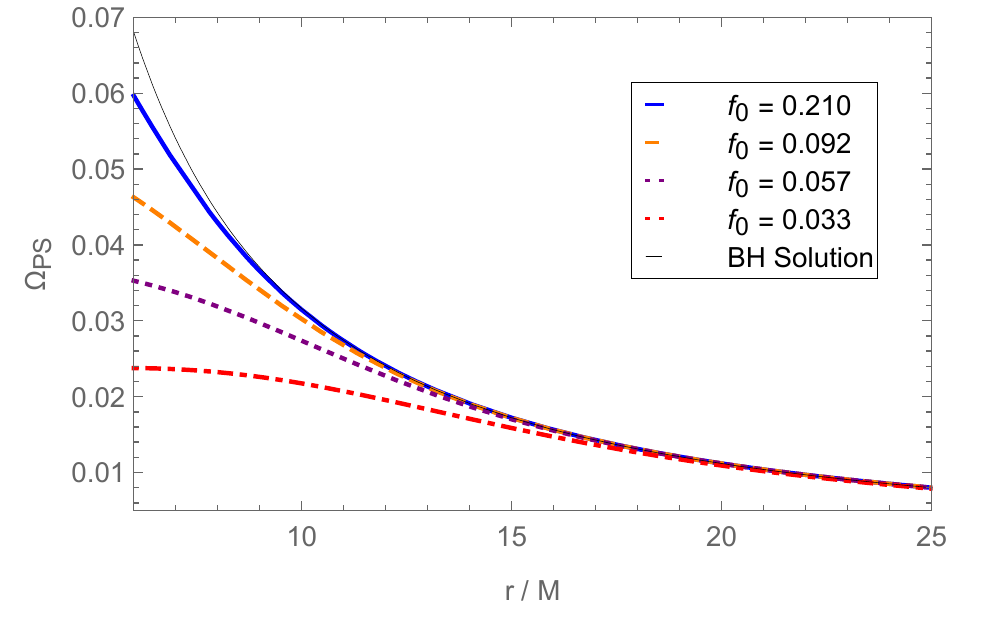}
\caption{Orbital velocity $\Omega$ from Eq.~\eqref{orbvel} as function of the radial coordinate $r$ normalized to the total mass $M$. \textbf{Top panel:} results for the four boson stars considered in Table \ref{tab:bsparam}; \textbf{Bottom panel:} results four Proca stars considered in Table \ref{tab:proca}. The thin black line represents the orbital velocity of the Schwarzschild solution. An increase in the central density $\gamma_0$ and $f_0$ leads to orbital velocities closer to that of a Schwarzschild solution.}
\label{fig:orbitvel}
\end{figure}

In Fig.\ref{fig:orbitvel} we plot the orbital velocities $\Omega$ for the eight solutions presented in Tables \ref{tab:bsparam} and \ref{tab:proca}.  It is clear that as we increase the central density of the boson and Proca star, i.e., as we increase $\gamma_0$ and $f_0$ respectively, the orbital velocities become closer to their BH counterparts. Furthermore, an increase in the orbital radius for any given solution with a specified central density also leads to a decrease in the differences between the bosonic star solution and the BH spacetime. An analysis of the orbital periods for each of the cases in comparison with the Schwarzschild case is provided in Appendix \ref{app:periods}.

\begin{figure}[b]
\includegraphics[scale=0.7]{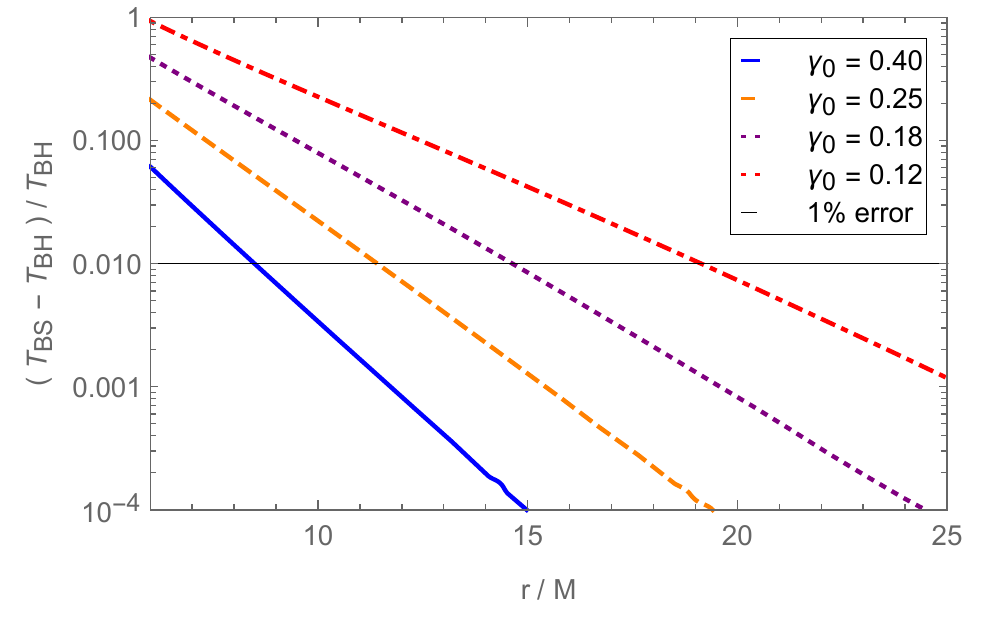}
\includegraphics[scale=0.7]{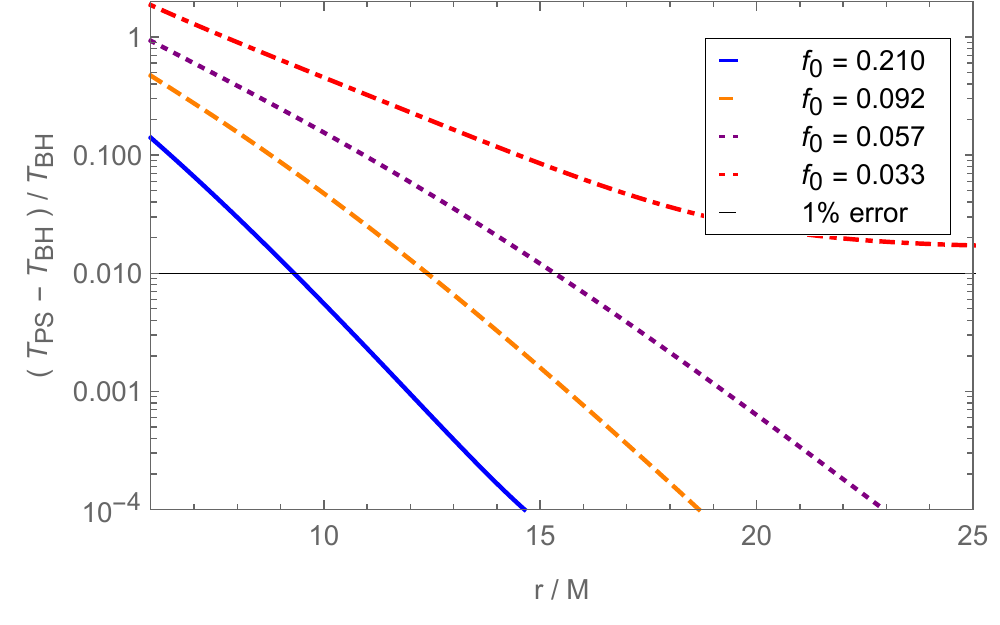}
\caption{Relative error of the orbital period $T=2\pi/\Omega$ as functions of the radial coordinate $r$ normalized to the total mass $M$. \textbf{Top panel:} results for the four boson stars considered in Table \ref{tab:bsparam}; \textbf{Bottom panel:} results for the four Proca stars considered in Table \ref{tab:proca}. The thin black line represents the $1\%$ error level. The relative errors decrease with an increase in the central densities $\gamma_0$ and $f_0$}
\label{fig:orbitper}
\end{figure}

Even though the analysis of the orbital velocities provides useful insights on the problem, it is the orbital period that becomes the measurable quantity when it comes to the comparison with observational data. Thus, we shall also analyse the orbital periods for each of the spacetimes considered. In Fig.\ref{fig:orbitper}, we plot the relative difference between the orbital periods of the boson stars $T_{BS}$ and Proca stars $T_{PS}$ in comparison to the orbital period in a Schwarzschild spacetime $T_{BH}$. Similarly to what happens with the orbital velocity, the orbital periods approach $T_{BH}$ for large radii, the difference eventually being smaller than the experimental uncertainties can cover.  

\section{Secondary image and critical orbital radius}
\label{app:angles}

\begin{figure}
    \centering
    \includegraphics[scale=0.8]{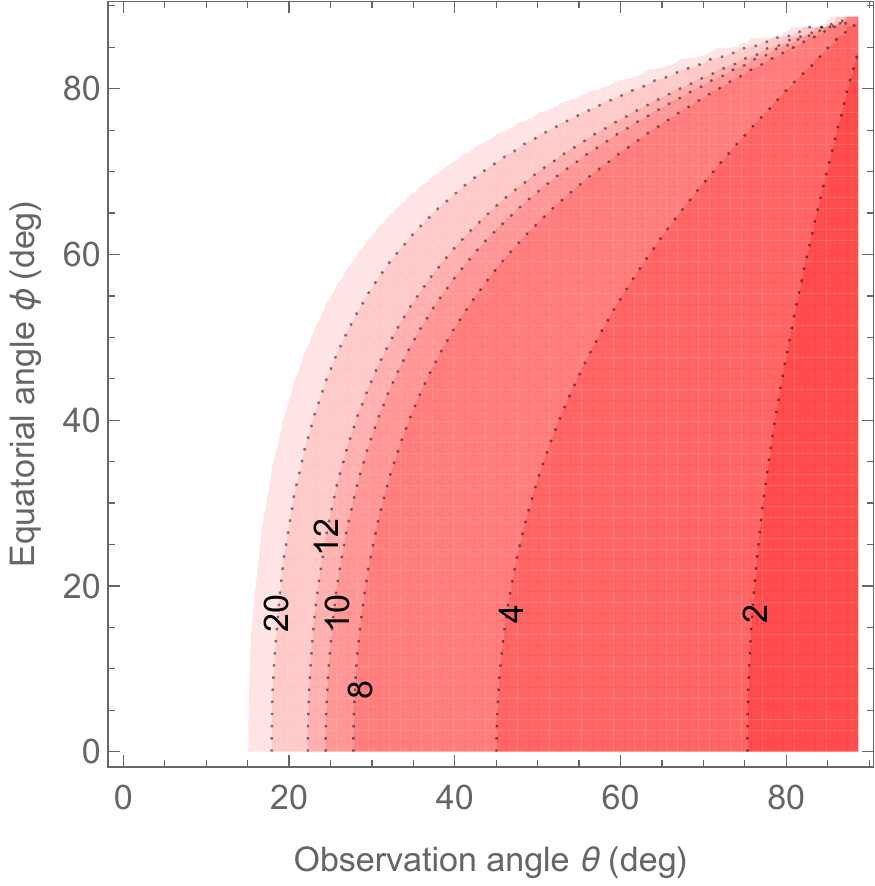}
    \includegraphics[scale=0.8]{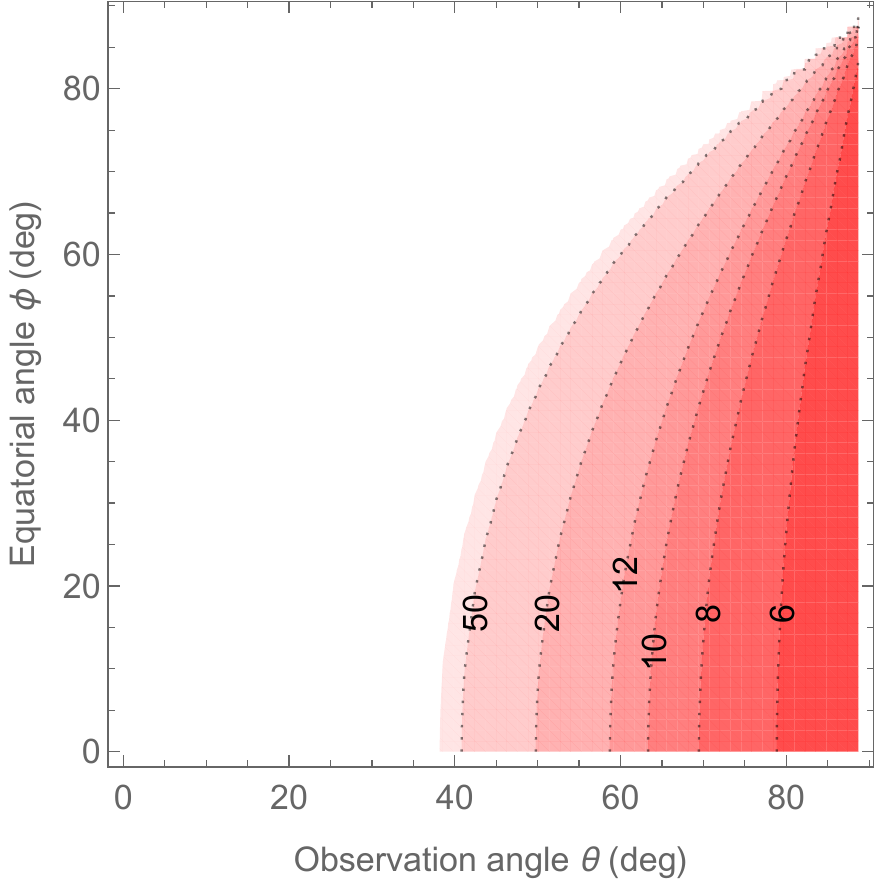}
    \caption{Minimum orbital radius necessary for a secondary image to appear in the screen of the observer (and, consequently, produce the shifting effect in the centroid) as a function of the observation angle of the observer $\theta$ and the equatorial angle of the source $\phi$, where $\phi=0$ was defined as the direction opposite to the observer, for the BSC4 (\textbf{top panel}) and the PSC4 (\textbf{bottom panel}) configurations.}
    \label{fig:criticals}
\end{figure}

In Sec.\ref{sec:lensing} we have stated, following Fig.\ref{fig:congruences}, that it is expected that for the same central massive object a secondary image might be absent if the observation is done at a low inclination but it might appear as one increases the observation angle. In this section, we aim to provide more details regarding this issue. In Fig.\ref{fig:criticals}, we plot the critical orbital radius (i.e., the minimum orbital radius necessary for the secondary image to appear in the screen of the observer) as a function of the observation angle $\theta$ (horizontal axis) and the equatorial angle $\phi$ (vertical axis), where we have defined $\phi=0$ when the hotspot is behind the central object as seen from the observer, for the most compact comfigurations of boson and Proca stars, i.e., BSC4 and PSC4. In these figures, the spiked boundaries near the leftmost contour correspond to numerical resolution limitations. 

Figure \ref{fig:criticals} explains the results previously obtained for the behaviour of the centroid of the flux: for the BSC4 configuration at an observation angle of $\theta=20º$ the effect is only visible for an orbital radius of $r_o=20M$, whereas if the observation inclination is increased to $\theta=50º$ the effect is visible for all the orbital radii considered in Fig. \ref{fig:centroid_C4}. For the PSC4 configuration, the effect is absent for all orbital radii considered when the observation angle is $\theta=20º$, if is present only for an orbital radius of $r_o=20M$ when $\theta=50º$, and it is present for all the orbital radii considered when $\theta=80º$. These results also show that the secondary image will be absent for the BSC4 configuration for an observation angle $\theta\sim 15º$ or smaller and for the PSC4 configuration for an observation angle $\theta\sim 38º$ or smaller, independently of the orbital radius. The existence of a minimal inclination angle for the secondary image to appear was already motivated previously in Fig. \ref{fig:congruences}.

Finally, Fig.\ref{fig:criticals} also provides information on the range of the equatorial angle for which the shift in the centroid is present. Consider e.g. the BSC4 configuration with an observation angle of $\theta=50^\circ$, from which Fig.\ref{fig:criticals} tells us that the effect will be visible for an equatorial angle up to $\phi~75^\circ$ to either side of the compact object for the orbital radius of $r_o=20M$, resulting in a range for the effect of about $150^\circ$, again consistent with the results of Fig.\ref{fig:centroid_C4}.
\section{Lensing data for full parameter space}
\label{app:integrated} 
\begin{figure*}[ht!]
    \includegraphics[width=1.8\columnwidth]{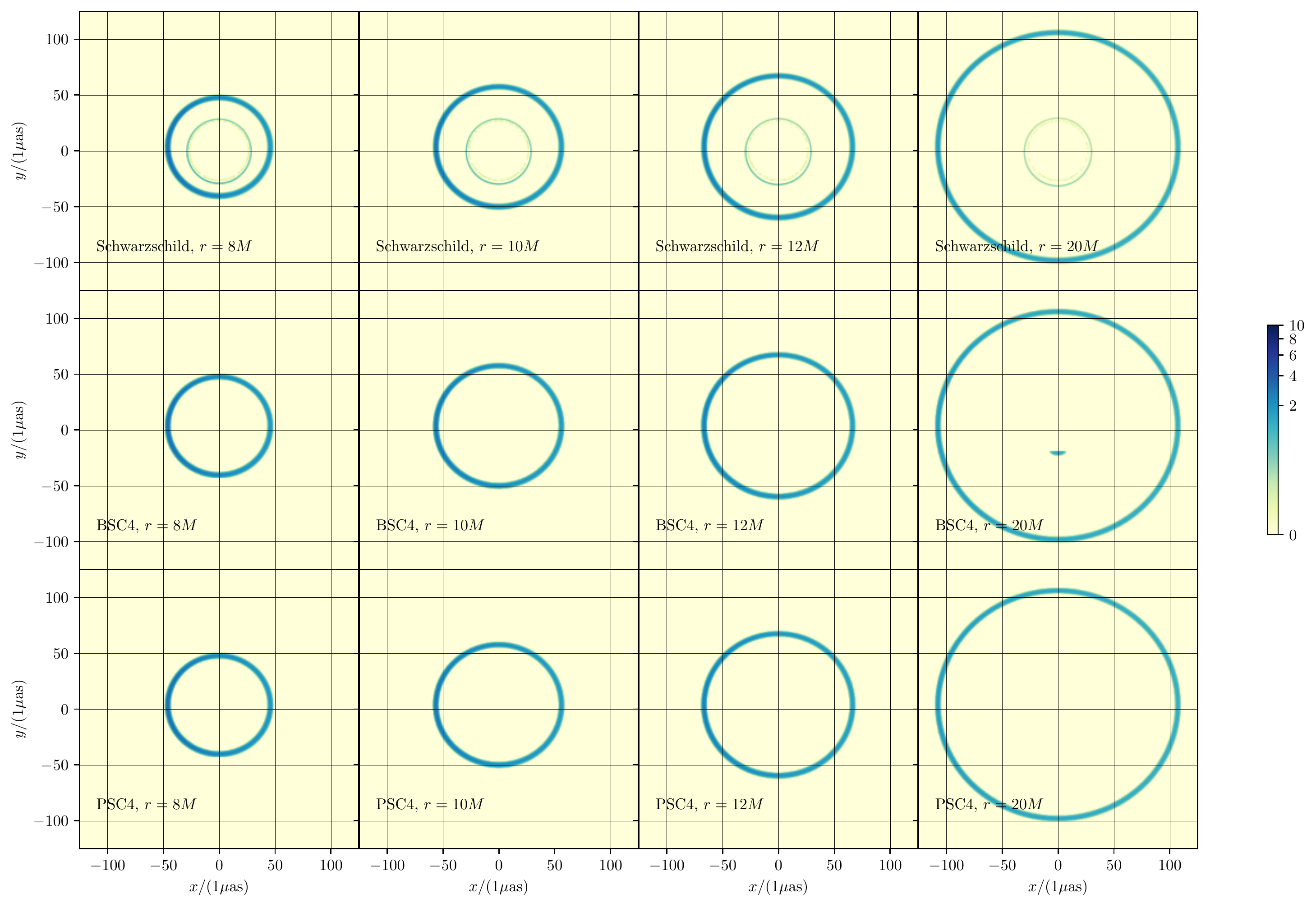}\\
    \includegraphics[width=1.8\columnwidth]{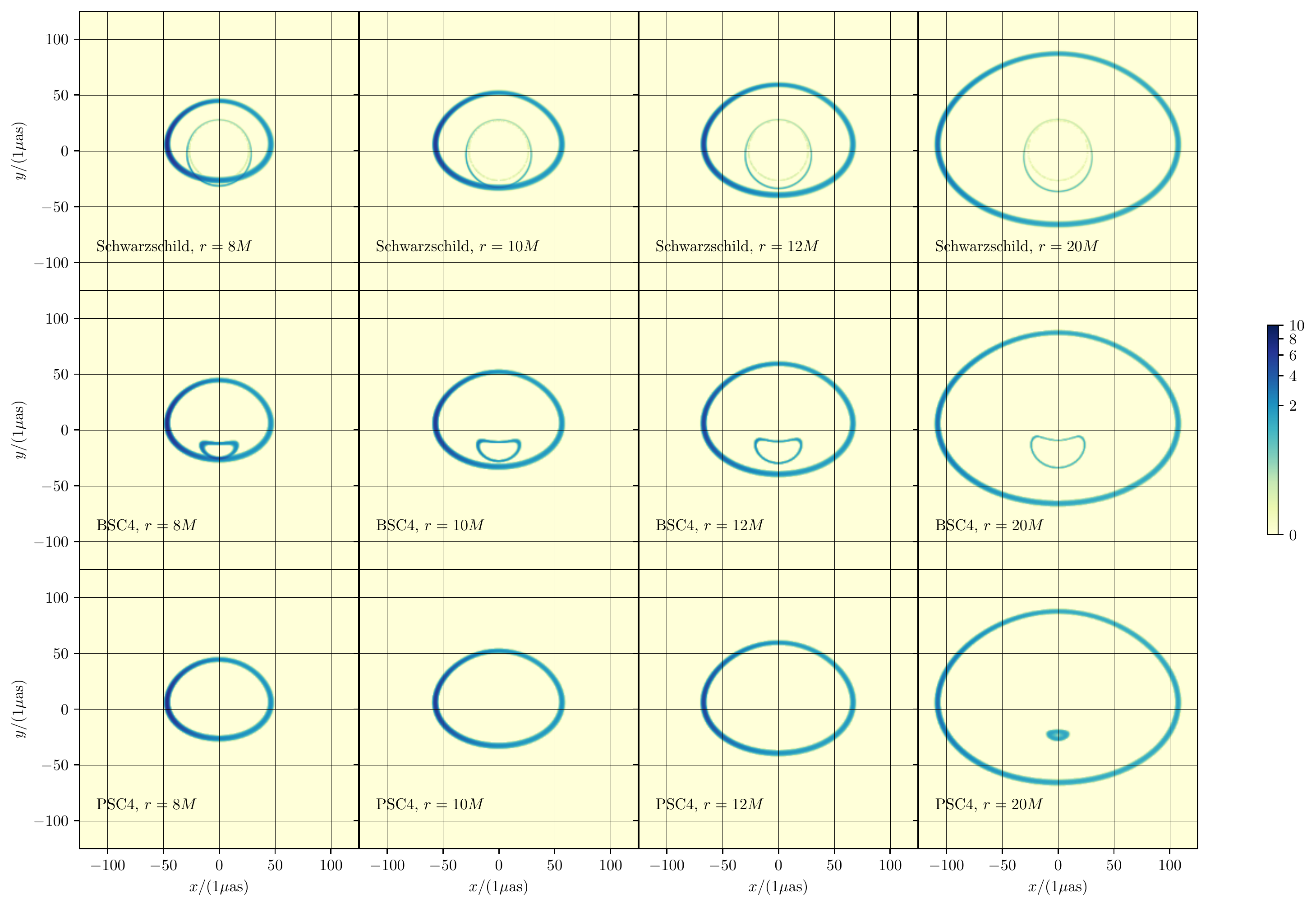}
	\caption{\label{fig:integrated_images_20} Time integrated images $\langle I\rangle_{lm}$ for a full orbit assuming a spherical spot. The observer inclination is $i=20^\circ$ (top 3 panel) and  $i=50^\circ$ (bottom 3 panels). The rows depict the compact object metrics (Schwarzschild, boson star and Proca star). The columns portray different orbital radius ($8M, 10 M, 12 M, 20M$).}
\end{figure*}

\begin{figure*}[ht!]
    \includegraphics[width=1.8\columnwidth]{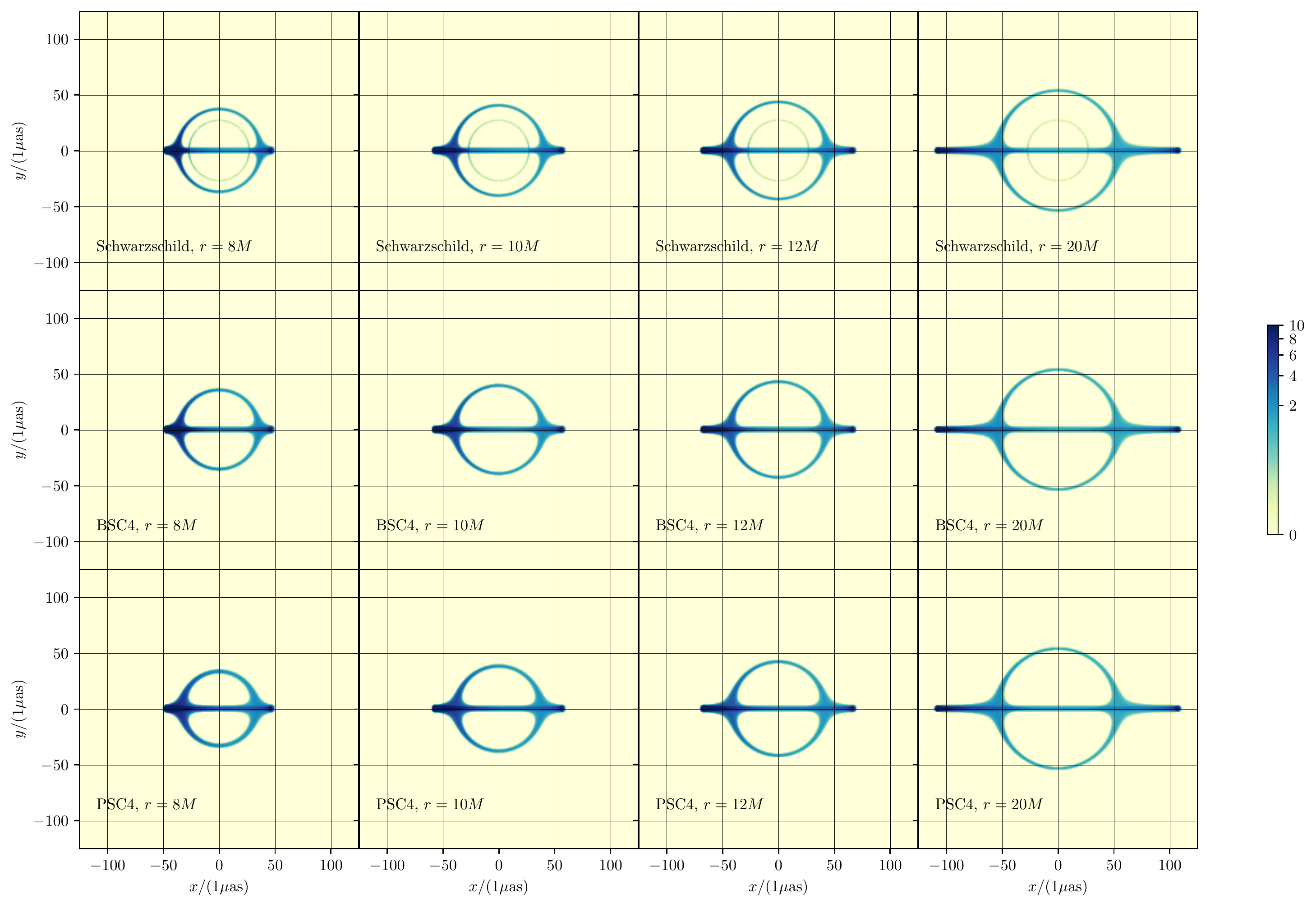}
	\caption{\label{fig:integrated_images_90} Time integrated images $\langle I\rangle_{lm}$ for an observer inclination of $i=90^\circ$ (see Figure~\ref{fig:integrated_images_20} for details).  }
\end{figure*}

\begin{figure*}[ht!]
    \includegraphics[width=2\columnwidth]{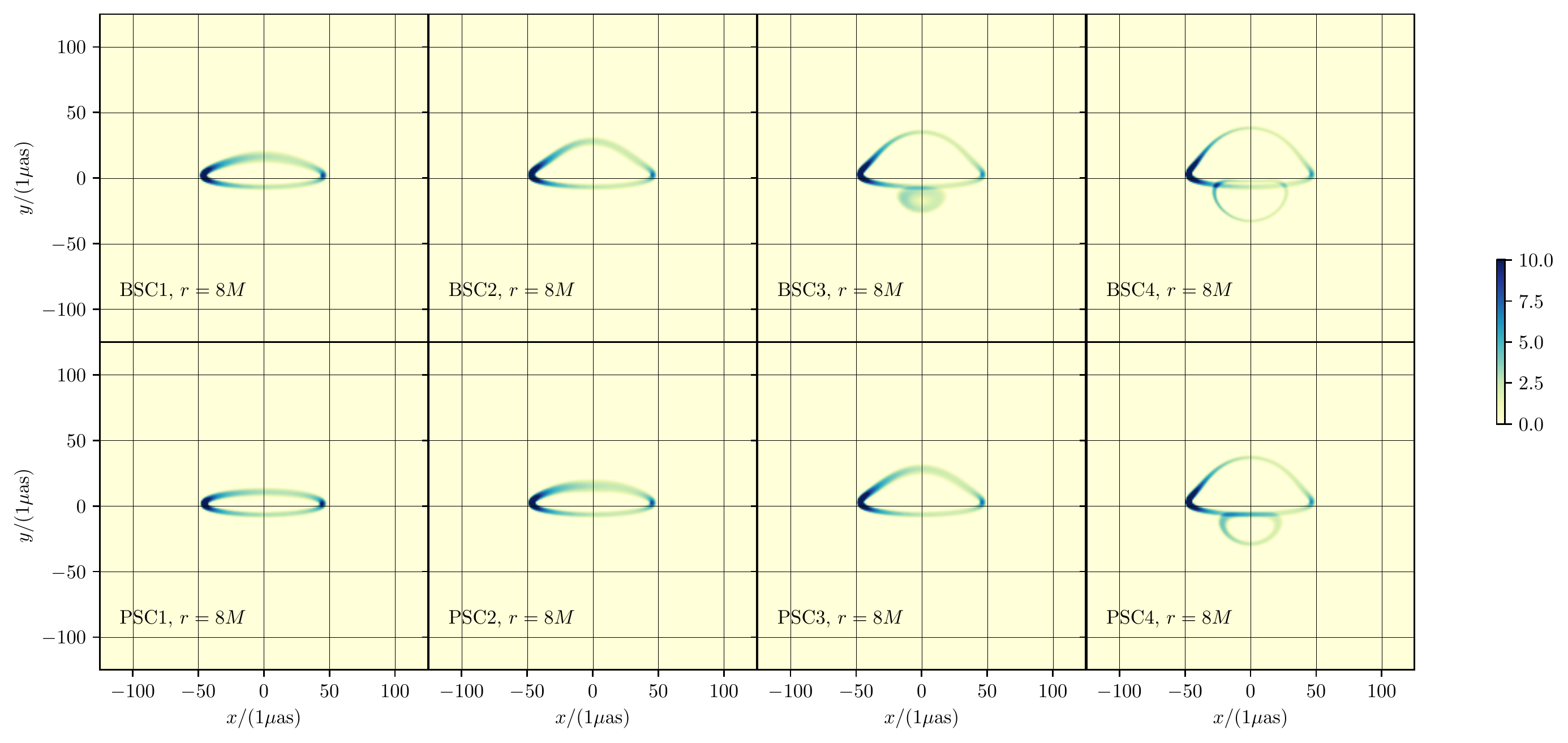}
	\caption{\label{fig:integrated_images_compactness} Time integrated images $\langle I\rangle_{lm}$ for an observer inclination of $i=80^\circ$ (see Figure~\ref{fig:integrated_images_20} for details), for boson and Proca stars of increasing compactness (cf. Tables~\ref{tab:bsparam} \& \ref{tab:proca}). The orbital radius is $r=8M$. As the compactness increases the lensing is stronger due to a larger mass of the compact objet inside the orbit.}
\end{figure*}

\begin{figure*}[h]
	\includegraphics[width=1.55\columnwidth]{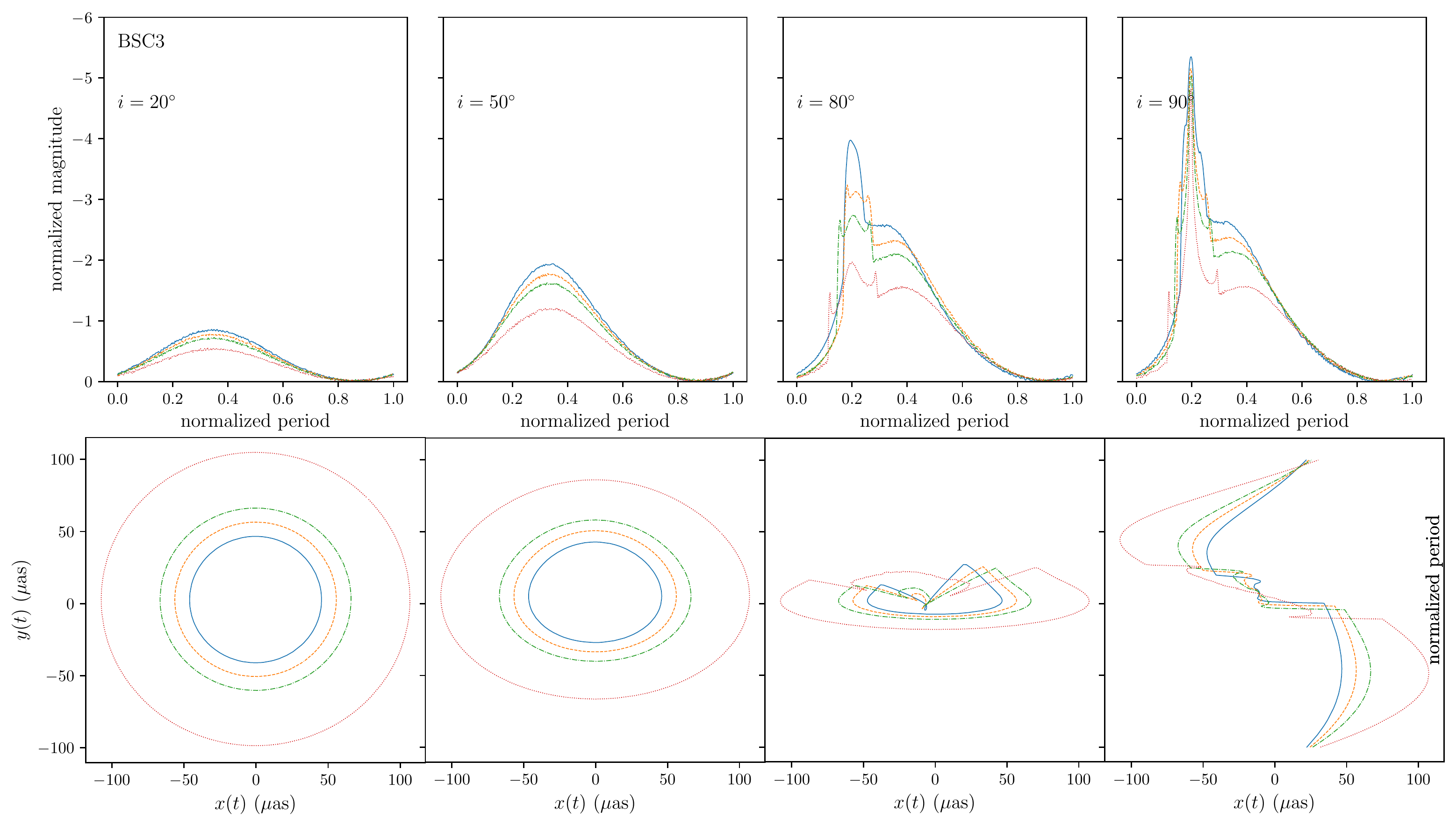}\\
	\includegraphics[width=1.55\columnwidth]{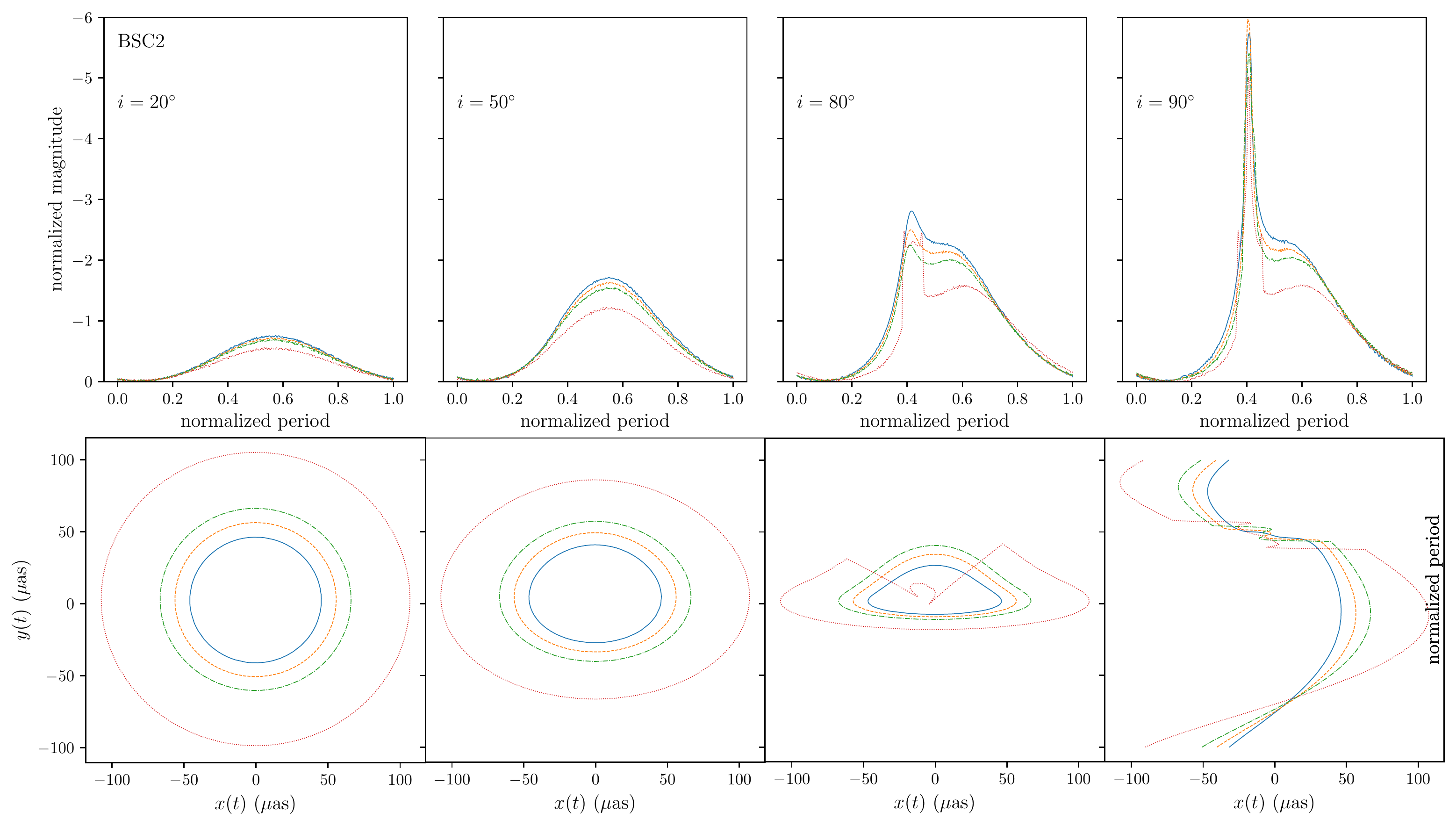}\\
	\includegraphics[width=1.55\columnwidth]{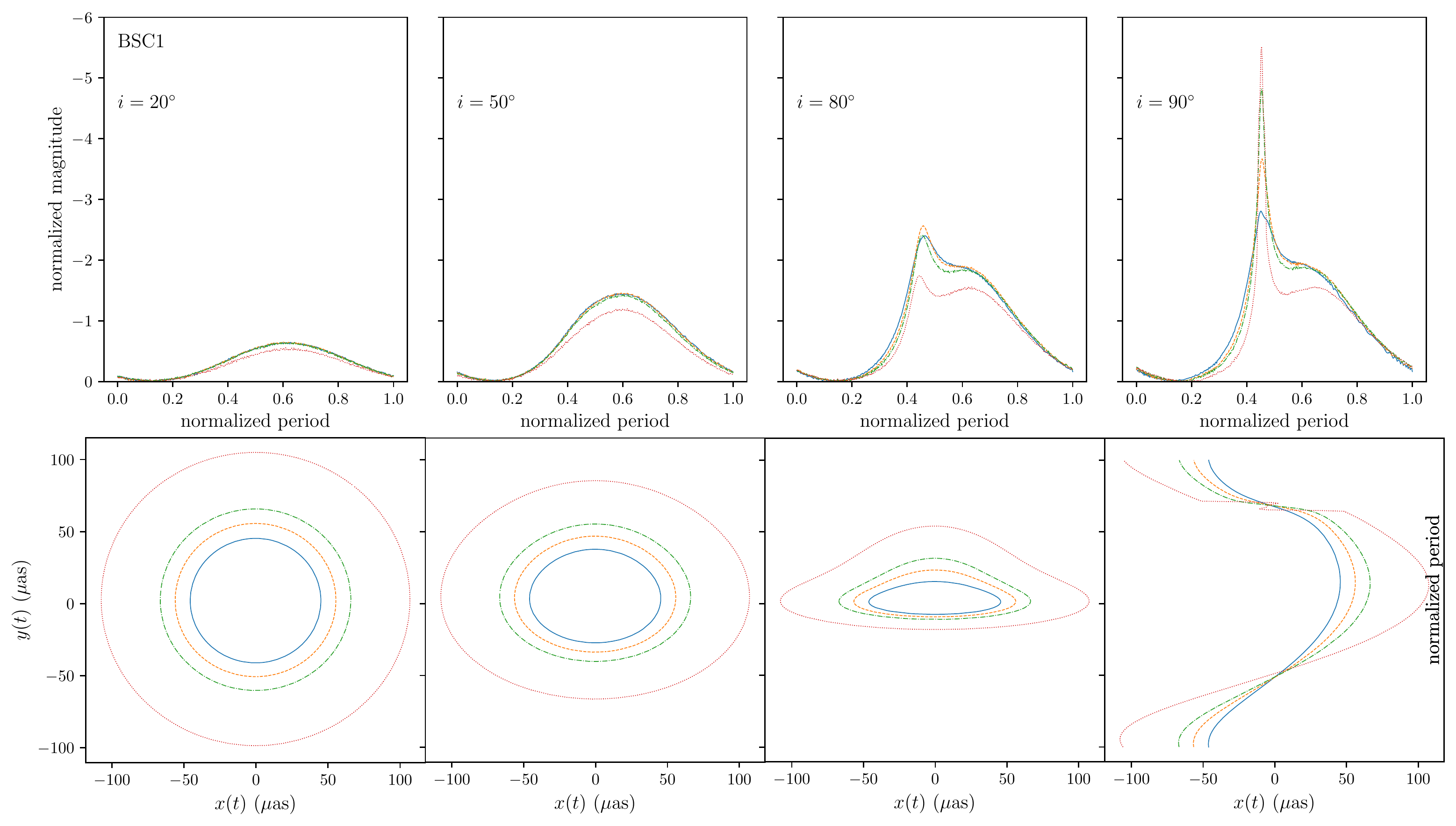}\\
	\caption{\label{fig:centroid_BC1-3} Temporal magnitude $m_k$ and temporal centroid $\vec{c}_k$ for the boson stars (cf. Table~\ref{tab:bsparam}). For $i=90^\circ$ centroid, the vertical axis is time.}
\end{figure*}

\begin{figure*}[h]
	\includegraphics[width=1.55\columnwidth]{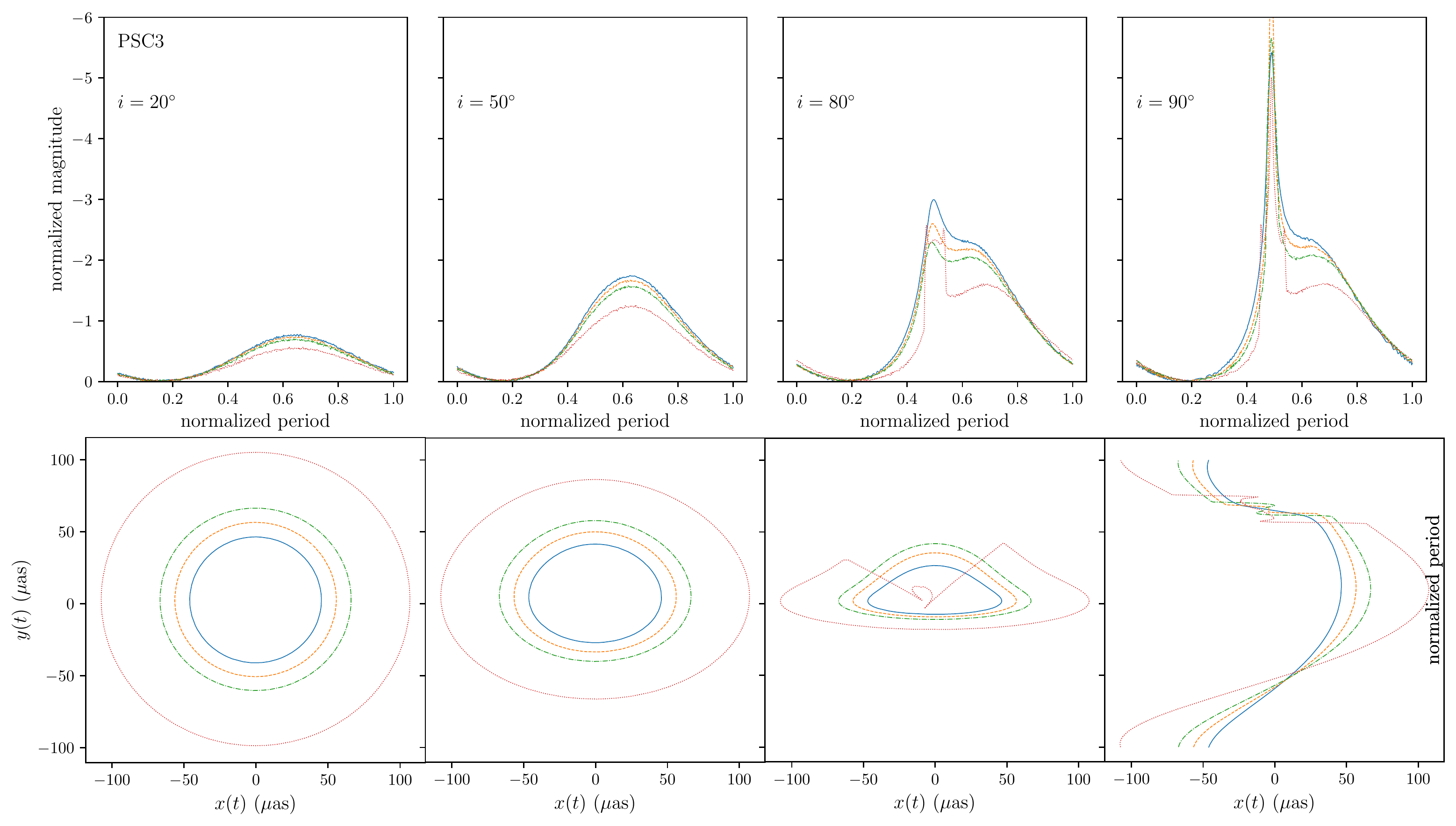}\\
	\includegraphics[width=1.55\columnwidth]{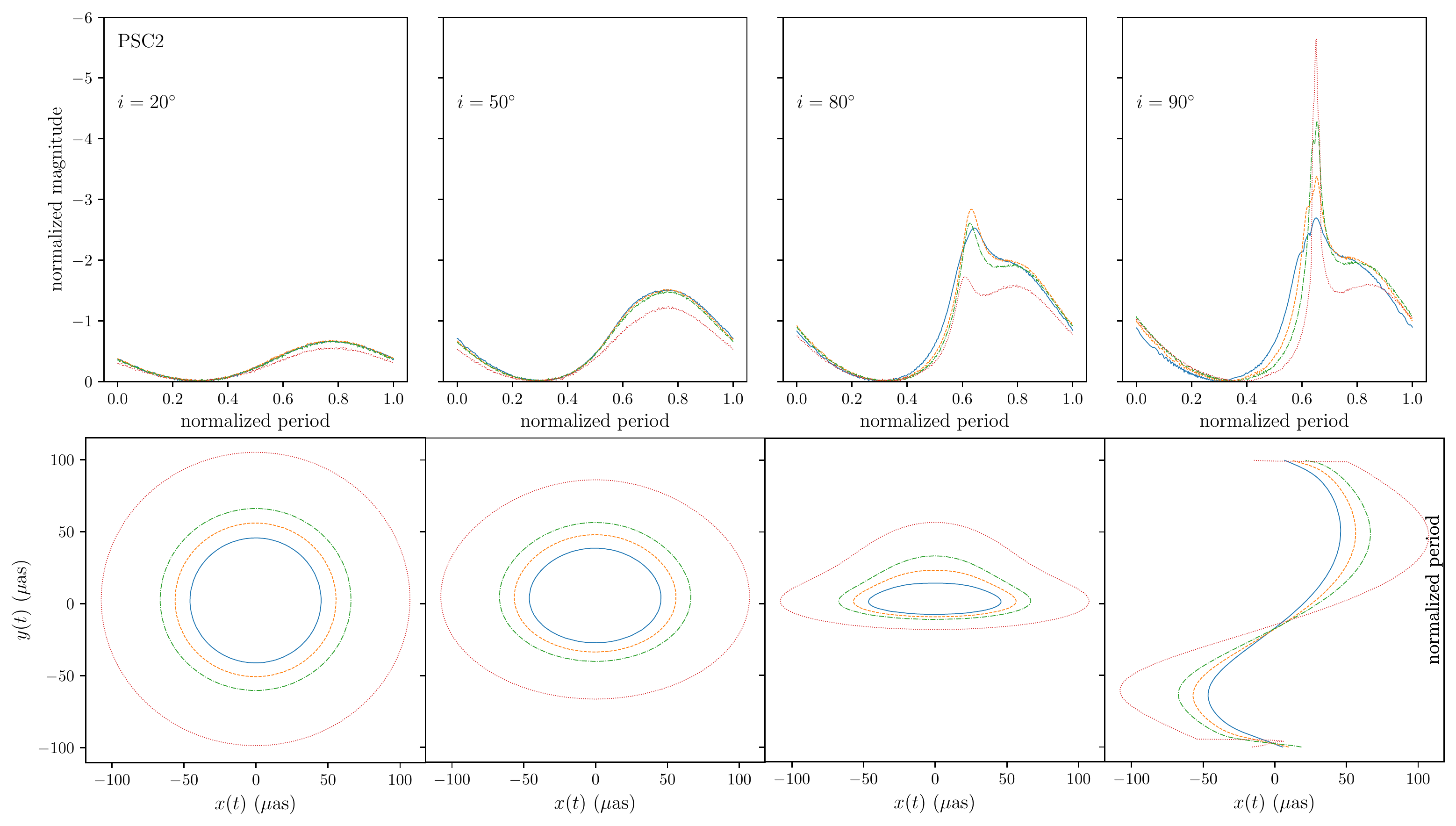}\\
	\includegraphics[width=1.55\columnwidth]{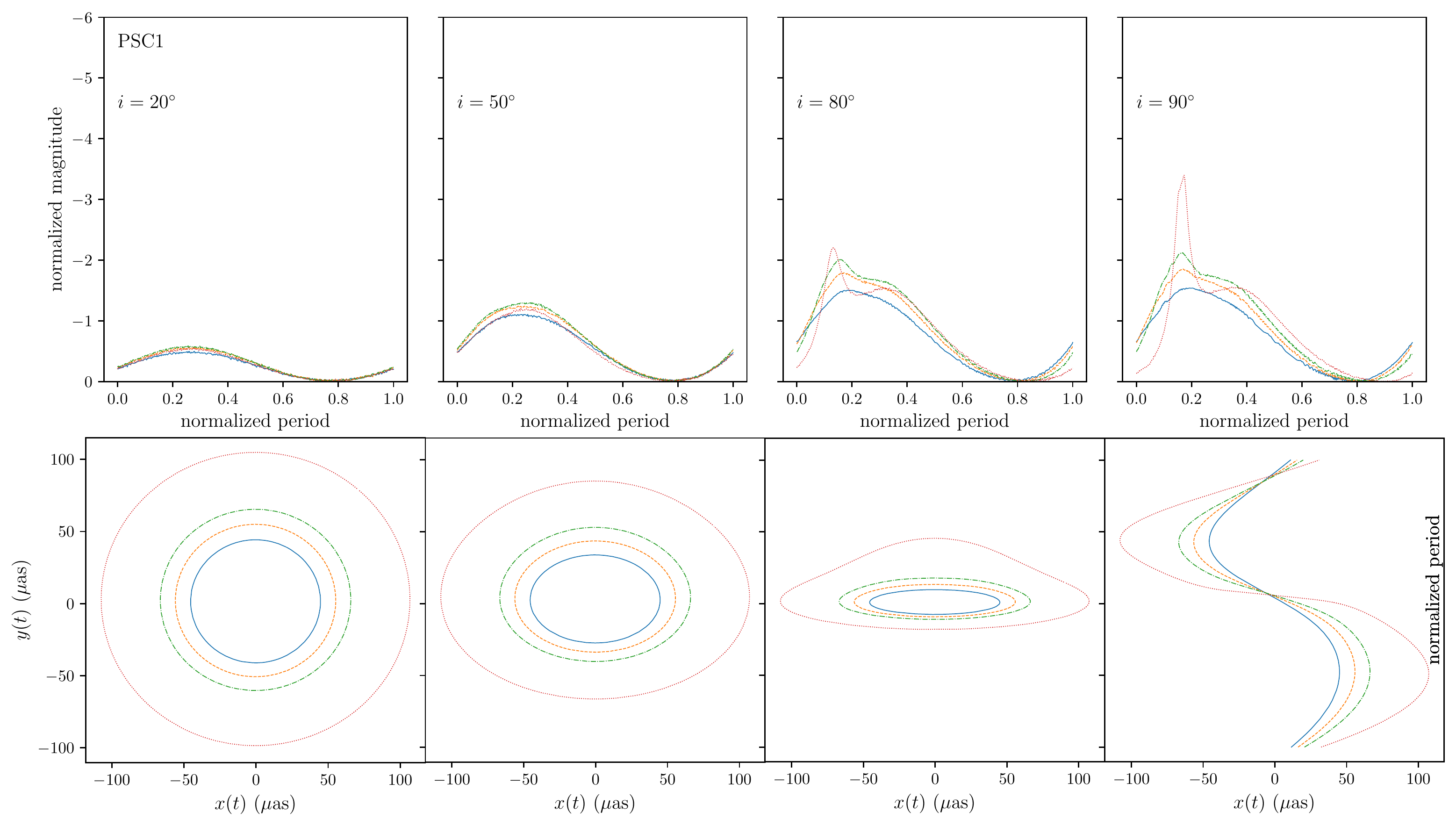}\\
	\caption{\label{fig:centroid_PC1-3}Temporal magnitude  $m_k$ and temporal centroid $\vec{c}_k$ for the Proca stars (cf. Table~\ref{tab:proca}). For $i=90^\circ$ centroid, the vertical axis is time.}
\end{figure*}

Here we extend the partial results of Sec.~\ref{sec:lensing} to the full parameter space. Figs.~\ref{fig:integrated_images_20}-\ref{fig:integrated_images_90} present time integrated images for the angles $i=20^\circ$, $50^\circ$ and $90^\circ$, for spot orbits $r=8M, 10 M, 12 M, 20M$, for the most compact objects. In Fig.~\ref{fig:integrated_images_compactness} the spot orbit is fixed $r=8M$ but the boson and Proca compactness is varied.  Fig.~\ref{fig:centroid_BC1-3} presents temporal fluxes and centroids for boson stars of compactness C1-C3 and Fig.~\ref{fig:centroid_PC1-3} for Proca stars.

%
%

\end{document}